\documentclass[11pt]{article}
\usepackage{mathrsfs}
\usepackage{epic, eepic}
\usepackage{subfigure}
\usepackage{amsfonts, amssymb, amsmath, graphicx, epsfig, lscape, capt-of}
\usepackage{url}
\usepackage{ifpdf}
\usepackage{ulem}
\def\inst#1{$^{#1}$}

\newtheorem{theorem}{Theorem}[section]

\newtheorem{definition}[theorem]{Definition}

\makeatletter

\begin{document}%

%%%%%%%%%%%%%%%%%%%%%%

\title{A new measure for community structures through indirect social connections}

\author{%
Roy Cerqueti \thanks{Corresponding author.}\inst{1}\and Giovanna
Ferraro\inst{2} \and Antonio Iovanella\inst{2}}
\date{}

\maketitle

\begin{center}
{\footnotesize
\inst{1} Department of Economics and Law\\
University of Macerata\\
Via Crescimbeni, 20 - 62100 Macerata, Italy\\
\texttt{roy.cerqueti@unimc.it}\\
\vspace{0.3cm} \inst{2} Department of Enterprise Engineering\\
University of Rome Tor Vergata\\
Via del Politecnico, 1 - 00133 Rome, Italy.\\
\texttt{giovanna.ferraro@uniroma2.it\\
antonio.iovanella@uniroma2.it}}
\end{center}

\begin{abstract}

Based on an expert systems approach, the issue of community
detection can be conceptualized as a clustering model for networks.
Building upon this further, community structure can be measured through a
clustering coefficient, which is generated from the number of
existing triangles around the nodes over the number of triangles
that can be hypothetically constructed. This paper provides a new
definition of the clustering coefficient for weighted networks under
a generalized definition of triangles. Specifically, a novel concept
of triangles is introduced, based on the assumption that, should the
aggregate weight of two arcs be strong enough, a link between
the uncommon nodes can be induced. Beyond the intuitive meaning of such
generalized triangles in the social context, we also explore the
usefulness of them for gaining insights into the topological structure
of the underlying network. Empirical experiments on the standard
networks of 500 commercial US airports and on the nervous system of
the Caenorhabditis elegans support the theoretical framework and
allow a comparison between our proposal and the standard definition
of clustering coefficient.

\vspace{5 mm} Keywords: complex networks; local cohesiveness,
clustering coefficient; generalized triangles.
\end{abstract}

%%%%%%%%%%%%%%%%%%%%%%
\section{Introduction}

Networks represent an effective methodological device for modeling
the main features of several complex systems~\cite{AB, New}. This paper builds on such a premise by focusing on the tendency of nodes in
a network to cluster, i.e. the link formation
between neighboring vertices~\cite{WS} leading to the identification
of the local groups cohesiveness. Such a theme is of paramount
relevance in that it allows one to assess the community structure of a
group of interconnected units~\cite{BDIF}. In this respect, we are in accord with Liu and Juan Ban~\cite{liu}, who
state that, in agreement with the expert systems perspective, the
problem of community detection can be dealt with as a clustering model
for networks. This explains also why community detection is nowadays
at the core of most discourse surrounding social networks (see e.g.~\cite{AGST, BPRC, dutch, ESA1}).

One of the most acknowledged and employed measures for assessing the
tendency of vertices to cluster is the {\it local  cluster
coefficient}~\cite{WS}. Such a quantity has
been extensively studied by several authors and applied in different
networks~\cite{Ops, WF, WS, ZWXMDF}. It captures the degree of social
embeddedness of the nodes in a network and is based on local density~\cite{SV}.
Indeed, especially in social
networks, vertices tend to create tightly knit groups that are
characterized by a relatively high density of links~\cite{Scott}.

The clustering coefficient assesses the connectivity of node
neighborhoods; a node having a high value of clustering coefficient
tends to be directly connected with well-established communities of
nodes~\cite{rosannafattore, CP}. Clustering coefficient is relevant
when determining the small-world property of a network~\cite{HG} and
can be considered as an index of the redundancy of a node~\cite{BSP,
LNP}. In the context of weighted networks, the clustering
coefficient has been analyzed in Grinrod~\cite{Grin}, Onnela et
al.~\cite{OCKKK, OSKK}, Barrat et al.~\cite{BBPV}, Zhang and
Horvath~\cite{ZH} and Opsahl and Panzarasa~\cite{OP}, as reported in
Section~\ref{tb}.

The weighted framework is of paramount relevance, in that the
analysis of the weights along the edges and their correlations is
able to provide a description of the hierarchical and structural
organization of the systems. This is evident if we consider, as an
example, a network in which the weights of all links forming
triangles of interconnected vertices are extremely small. In this
case, even for a large clustering coefficient, these triangles play
a minimal role in the network dynamics and organization, and the
clustering features are certainly overestimated by a simple
structural analysis~\cite{BBPV}. Also, vertices with high degree can
be attached to a majority of low-degree nodes whilst concentrating
the largest portion of their strength only on the vertices with high
degree. In this situation, the topology reveals a disassortative
characteristic of the network, whereas the system could be
considered assortative since the more relevant edges in terms of
weights are linked to the high-degree vertices~\cite{leungchau}.

Despite several measures being proposed for the local and global
clustering coefficients, they are all only able to capture the clustering of ego
networks or the overall statistics regarding the network~\cite{BKJ,
PEF}. An ego is a focal individual and the ego network is composed
of the nodes directly connected to him (also called alters) and the
links among him and others (see e.g.~\cite{biswas}).

Thus, in this paper we are interested in two
relevant cases. In the first, the ego is connected to two alters not
mutually connected and we aim to understand if the strength of the
connections with the ego is strong enough to induce a certain
level of interaction as can be found when they are connected.

In the second, the alter of an alter is not directly connected to the ego.
Also in this case, we advance the proposal that the strength of the
existing connections induces interactions between the ego and the
alter of the alter.

It is worth noting that all the considered aspects can be
interpreted in the context of link formation as reasonable premises.
Link prediction is relevant in that it attempts to estimate the
likelihood of a link existing between two vertices based on
observed links and the attributes of nodes~\cite{AA}.
Such a prediction can be used to analyze a network to suggest
promising interactions or collaborations that have not yet been
identified, or is related to the problem of inferring missing or
additional links that, while not directly visible, are likely to
exist~\cite{LNK, LZ}.

The specific aim of this paper is to introduce a novel definition
of a generalized clustering coefficient by including also the
triples of the two cases presented above. In so doing, our concept
of community captures the weighted network's propensity for close
triples. Moreover, this measure is also useful for predicting the
fictitious links that may appear in the future of evolving networks.

Our generalized clustering coefficient has a further relevant
property: it assumes unitary value not
only when the graph is a clique, but in a number of different situations.
Specifically, the community
structure of the network is intended to include also the realistic
cases of the presence of indirect connections among two agents
induced by their strong links with a third node.

The ground of our study is quite intuitive. Indeed, in the context
of community structure of weighted networks, there is evidence that
strong enough connections among two individuals are prone to creating
triangles among their neighborhood. Formally, this means that it is
possible to introduce a threshold for stating when the weight of a
link can be defined as strong enough. We reasonably take that the
larger the threshold, the stronger the link.

In this respect, as we will see below in the formalization of our
setting, null thresholds mean no constraints -- and all the
two-sided figures can be viewed as triples -- while a large value of
the thresholds is associated to very restrictive constraints -- and
a small number of two-sided figures will be accepted as triples.

It is very important to note that the case of zero thresholds
gives further insights into the topological structure of the unweighted
graph associated to the network. We direct the reader to the
empirical analysis section for an intuitive explanation of this
point. In this regard, we have also implemented a comparison between
our definition and the standard clustering coefficient for weighted
networks.

Based on such a perspective, this paper also implements a wide
computational analysis to explore the reaction of the proposed
clustering coefficient to threshold variations.

The paper is structured as follows. Section 2 outlines the
motivations -- based also on real-world applications -- behind the
present study and the novel definition of clustering coefficient.
Such a motivating discussion is proposed before the formal
definition to immediately convince the reader of the usefulness of
the presented scientific proposal. For some more formal insights on
the generalized clustering coefficient and on the generalized
triangles, please refer to Section 5, where a detailed discussion of
definitions and concepts is carried out. Section 3 is devoted to the
outline of certain relevant preliminaries and the employed notations
about the graph theory. Section 4 contains a review of the
literature on the clustering coefficient in both cases of weighted
and unweighted networks. Section 5 introduces and discusses the
proposed definition of generalized clustering coefficient and
generalized triples, along with the related interpretation. Section
6 focuses on the computational experience of two empirical networks:
the network among the 500 commercial airports in the United States
and the nervous system of the nematode Caenorhabditis elegans. The
final section offers some conclusive remarks and outlines directions
for future research.

%%%%%%%%%%%
\section{Motivation for the generalized clustering coefficient and real-world applications}

One of the major fields of study in the empirical investigation of
networks is the uncovering of subgroups of nodes according to a
given criteria. Such subgroups, or communities, are interesting
since they can help to understand a wide variety of possible group
organizations, and they occur in networks in biology, computer science,
economics, politics and more~\cite{Fort, GN, WC}. Recently,
community discovery has been used in social media, such as in~\cite{XLC},
where authors propose a community-aware approach to constructing
resource profiles via social filtering, in~\cite{ZTHHZ}, where
communities are discovered from social media by low-rank matrix
recovery, and in~\cite{ESA1}, where communities are studied by means
of the network's internal structural properties.

\begin{figure}[tbp]
\begin{center}
\includegraphics[scale=.45]{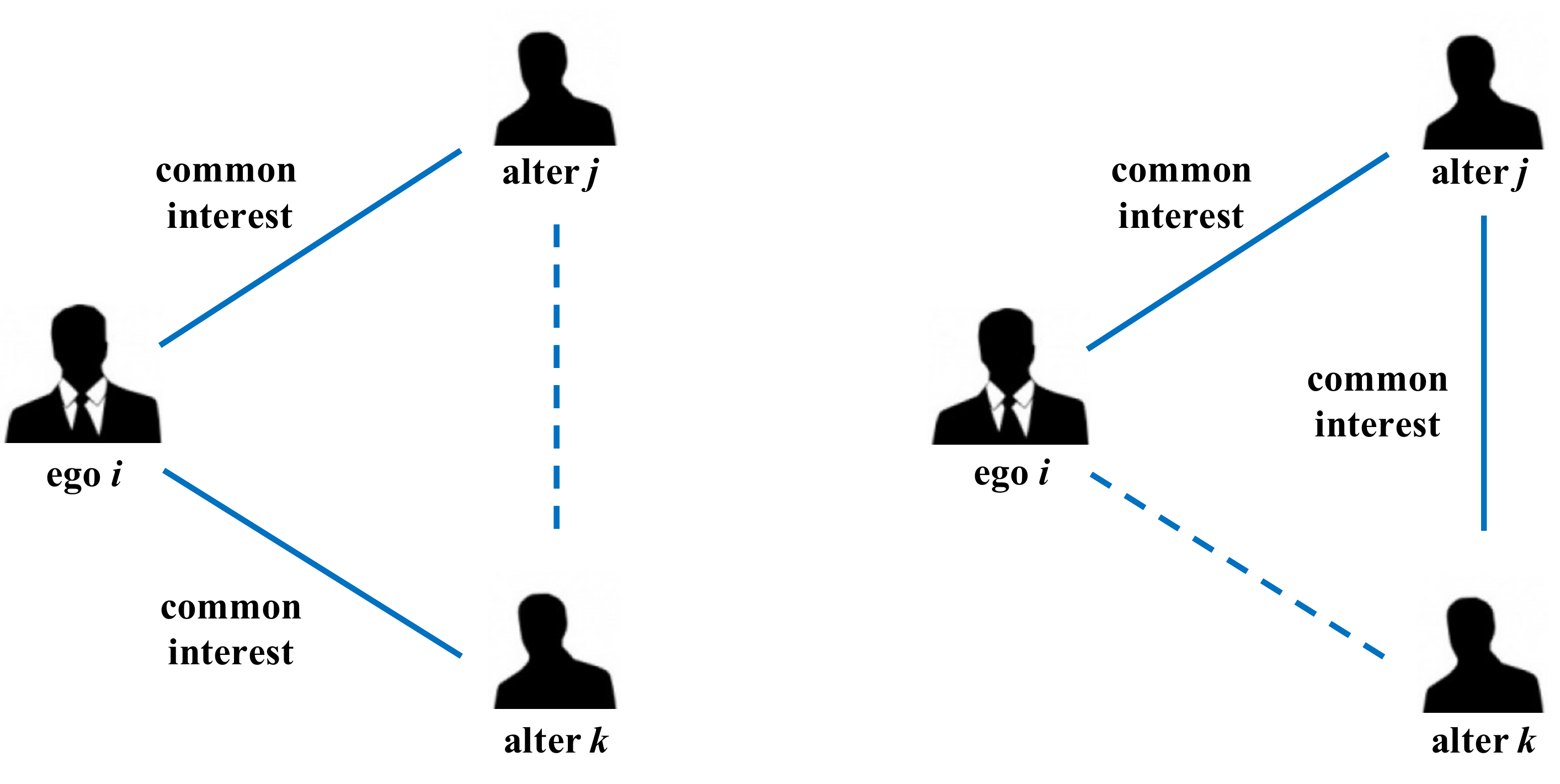}
\caption{Two different schemes of relationships based on mutual common interests,
from an ego and its ego network (left) and from an ego and a path of
length 2 (right).}
\label{common}
\end{center}
\end{figure}

The behavior of nodes is often highly
influenced by the behavior of their neighbors or community
members~\cite{Fort, XLMLCR}. From this point of view, the clustering
coefficient is one of the main measures used to understand the
level of cohesion around a node.

The generalized concept of clustering coefficient presented
here describes community structures which are not established, but
are indirectly induced by strong cooperations among the formally
linked nodes. More specifically, the existence of a powerful link
between two nodes is assumed to be able to form the connection
between nodes that are disconnected but adjacent to the
considered ones.

In this respect, social motivations from the perspective of a single
node (ego) support the study of the proposed generalized clustering
coefficient. In particular, the knowledge for two alters of a common
ego increases the opportunities for them to meet since they could be
engaged in similar interests, which would ultimately provide a basis for them to trust one
another~\cite{EK}. Moreover, social psychology suggests that an ego has
an incentive to connect to
its alters in order to reduce its
isolation~\cite{Heider}. An ego might also be interested in
connecting to its neighbors for proximity reasons, taking into
consideration the shortest path distance between them~\cite{BKJ}.

Here, we also consider the possibility of having an indirect
connection of an ego with an alter of one of its alters. Therefore,
we consider two forms of engagements between an ego and various
alters, as shown in Figure~\ref{common}. More details on the figure
will be given in Section~\ref{implications}.

The perspective adopted here represents the basis for several
real-world cases. For example, in~\cite{HPKE}, affiliation networks
allow one to observe the connections among individuals indirectly,
i.e. not through directly observed social interactions, while
in~\cite{BKJ}, a new measure of the clustering coefficient is
proposed, with applications in the study of segregation and
homophily. Finally, in biology, gene expression data can be studied
with the weighted clustering coefficient in order to reveal
differences between normal and tumour networks~\cite{KH}.

It is also worth mentioning how the concepts of triples introduced
here could relate to the issue of link formation. Indeed, the
presence of a strong connection between two units would probably
induce cooperation also among the units connected with those
considered in the near future. In this regard, link creation
was studied from the perspective of the clustering coefficient. We
mention~\cite{KW}, where authors investigate the origins of
homophily and tie formation by means of triadic closures and
proximity, and~\cite{PEF}, where a new method for triple estimation is
presented.

%%%%%%%%%%%%%%%%%%%%%
\section{Preliminaries and notations about graph theory}

For the convenience of the reader, we shall now provide some preliminaries and notations.
The classical mathematical abstraction of a network is a graph $G = (V, E)$, where
$V$ is the set of $N$ nodes (or vertices) and $E$ is the set of $M$
links (or edges) stating the relationships among the nodes. We refer
to a node by an index $i$, meaning that we allow a one-to-one
correspondence between an index in $\{1, \dots, N\}$ and a node in
$V$. The set $E$ can be
conceptualized through the adjacency matrix $\mathbf{A}=(a_{ij})_{i,j=1,
\dots,N}$, whose generic element $a_{ij}$ is 1 if the link between
$i$ and $j$ exists and 0 otherwise. The graph is undirected when
$a_{ij}=a_{ji}$, for each $i,j=1, \dots, N$, and directed otherwise.
The degree $d_i$ of the node $i$ is a nonnegative integer
representing the number of links incident upon $i$.

In this paper, we examine weighted networks, and we refer to a weighted adjacency
matrix $\mathbf{W}$ whose elements $w_{ij} \geq 0$ represent the
weights on the link connecting nodes $i$ and $j$, with $i, j = 1,
\dots, N$. Clearly, $w_{ij}=0$ stands for absence of a link between
$i$ and $j$. Thus, $w_{ij}$ denotes the intensity of the
interactions between two nodes $i$ and $j$ and allows for the modeling of
the ties' strength of the observed system.

%%%%%%%%%%%%%%%%%%%%%%

\section{Literature review on the clustering
coefficient for unweighted and weighted networks}\label{tb}

\subsection{Unweighted networks}
The {\it local clustering coefficient} can be defined for any vertex
$i=1, \dots, N$ and captures the capacity of edge creations among
neighbors, i.e. the tendency in the network to create stable
groups~\cite{WS}. Thus, the cohesion around a vertex
$i$ is quantified by the local clustering coefficient $C_i$ defined
as the number of triangles $t_i$ in which vertex $i$ participates
normalized by the maximum possible number of such triangles:

\begin{equation}
\label{Ci} C_i = \frac{2t_i}{d_i(d_i-1)}.
\end{equation}

\noindent The local clustering coefficient quantifies how a node
takes part in a cohesive group. Therefore, $C_i = 0$ if none of the
neighbors of a node are connected and  $C_i = 1$ if all of the
neighbors are linked.

The value of the local clustering coefficient is influenced by the
nodes degrees. A node with several neighbors is likely to be
embedded in fewer closed triangles; hence, it has a smaller
local clustering coefficient when compared to a node linked to fewer
neighbors, where they are more likely to be clustered in triangles~\cite{barbook}.

The clustering coefficient for a given graph is computed in two
classical modes~\cite{New}. The first is the {\it averaged
clustering coefficient} $\overline{C}$, given as the average of all
the local clustering coefficients, while the second, called the {\it
global clustering coefficient} and denoted by $C_G$, is defined as
the ratio among three times the number of closed triangles in the
graph and the number of its triples, i.e. the number of 2-paths
among three nodes.
%
%\begin{equation}
%\label{Cg} C_G = \frac{3 \times triangles}{triplets}
%\end{equation}

Note that both $\overline{C}$ and $C_G$ assume values from $0$ to
$1$ and are equal to $1$ in case of a clique, i.e. a fully coupled
network. In real networks, the evidence shows that nodes are
inclined to cluster into densely connected groups~\cite{FI1, WC}
and the difficulty of comparing
the values of clustering nodes with different degrees makes the
average value of local clustering sensitive to the way in which
degrees are distributed across the whole network.

The quantities $\overline{C}$ and $C_G$ are specifically tailored to
unweighted networks, and they cannot be satisfactorily employed to describe
the community structure of the network in the presence of
weights on links and when arcs are of the direct type.

The next section is devoted to the analysis of the more general weighted
cases.

\subsection{Weighted networks}
In many real networks, connections are relevant not only in terms of
the classical binary state -- whether they exist or do not exist --
but also with regards to their strength which, for any node $i=1,
\dots, N$, is defined as:

\begin{equation}
\label{Si} s_i = \sum_{j=1}^N w_{ij}.
\end{equation}

The introduction of weights and strengths extends the study of the
macroscopic properties of the network by adding some forms of entity of connections and capability to the mere interactions.
In particular, the strength integrates information about the vertex
connectivity and the weights of its links~\cite{BBPV}. It
is considered a natural measure of the importance or centrality of a
vertex $i$. Indeed, the identification of the most central nodes
represents a major issue in network characterization~\cite{Free}.

In~\cite{BBPV}, Barrat et al. combine the topological information of the
network with the distribution of weights along links, and define the
weighted clustering coefficient for a node $i=1, \dots, N$ as
follows:

\begin{equation}
\label{CiB} \tilde{C}_{i,B} = \frac{1}{s_i(d_i-1)}\sum_{j,k\in V}
\frac{w_{ij} + w_{ik}}{2} a_{ij}a_{jk}a_{ik}.
\end{equation}

This coefficient is a quantity of the local cohesiveness, which
considers the importance of the clustered structure by taking into
account the intensity of the interactions found on the local
triangles. This measure counts, for each triangle created in the
neighborhood of the node $i$, the weight of the two related edges.
The authors refer not to the mere number of the triangles in the
neighborhood of a node but also to their total relative weight with
respect to the strength of the nodes.

The normalization factor $s_i(d_i-1)$ accounts for the strength
$s_i$ and the maximum possible number of triangles in which the node
$i$ may participate, and it ensures that $0 \leq \tilde{C}_{i,B}
\leq1$. The definition of $\tilde{C}_{i,B}$ recovers the topological
clustering coefficient in the case where $w_{ij}$ is constant, for
each $j$.

Therefore, the authors introduce the weighted clustering coefficient
averaged over all nodes of the network, say $C^W$, and over all
nodes with degree $d$, say $C^W(d)$. These measures offer global
information on the correlation between weights and topology by
comparing them with their topological analogs.

Note that $s_i = d_i(s_i/d_i) = d_i\langle w_i \rangle$, so $\tilde{C}_{i,B}$ can be written as:

\begin{equation}\label{CiB2}
\tilde{C}_{i,B} = \frac{1}{d_i(d_i-1)}\sum_{j,k\in V} \frac{w_{ij} +
w_{kj}}{2\langle w_i \rangle} a_{ij}a_{jk}a_{ik}
\end{equation}

\noindent where $\langle w_i \rangle = \sum_j w_{ij}/{d_i}$. In such
equation the contribution of each triangle is weighted by a ratio of
the average weight of the two adjacent links of the triangle to the
average weight $\langle w_i \rangle$.

Thus, $\tilde{C}_{i,B}$ compares the weights related with triangles
to the average weight of edges connected to the local node.

Zhang and Horvath~\cite{ZH} describe the weighted clustering
coefficient in the context of gene co-expression networks. Unlike
the unweighted clustering coefficient, the weighted clustering
coefficient is not inversely related to the connectivity. Authors
show a model that reveals how an inverse relationship between the
clustering coefficient and connectivity occurs from hard
thresholding. In formula:

\begin{equation}
\label{CiZ} \tilde{C}_{i,Z} = \frac{\sum_{j,k\in V}
\hat{w}_{ij}\hat{w}_{jk}\hat{w}_{ik}}{(\sum_{k\in V} \hat{w}_{ik})^2
- \sum_k \hat{w}_{ik}^2}
\end{equation}

\noindent where the weights have been normalized by $\max(w)$. The
number of triangles around the node $i$ can be written in terms of
the adjacency matrix elements as $t_i = 1/2 \sum_{i,k\in V}
a_{ij}a_{jk}a_{ik}$ and the numerator of the above equation is a
weighted generalization of the formula. The denominator has been
selected by considering the upper bound of the numerator, ensuring
$\tilde{C}_{i,Z} \in [0,1]$ . The equation (\ref{CiZ}) can be
written as:

\begin{equation}
\tilde{C}_{i,Z} = \frac{\sum_{j,k\in V}
\hat{w}_{ij}\hat{w}_{jk}\hat{w}_{ik}}{\sum_{j,k \in V:j \ne k}
\hat{w}_{ij}\hat{w}_{ik}}
\end{equation}

In Grindrod~\cite{Grin}, a similar definition has been shown; indeed, the
edge weights are considered as probabilities such that in an
ensemble of networks, $i$ and $j$ are linked with probability
$\hat{w}_{ij}$. Finally, Holme et al.~\cite{HPKE} discuss the definition
of weights and express a redefined weighted clustering coefficient
as:

\begin{equation}
\tilde{C}_{i,H} = \frac{\sum_{j,k\in V} w_{ij}w_{jk}w_{ik}}{\max(w)
\sum_{j, k\in V} w_{ij}w_{ik}} =
\frac{\mathbf{W}^3}{(\mathbf{W}\mathbf{W}_{max}\mathbf{W})_{ii}}
\end{equation}

\noindent where $\mathbf{W}_{max}$ indicates a matrix where each
entry equals $\max(w)$. This equation seems similar to those in~\cite{ZH},
 though, $j \ne k$ is not required in the
denominator sum.

Onnela et al.~\cite{OCKKK, OSKK} refer to the notion of motif, defining it
as a set (ensemble) of topologically equivalent subgraphs of a
network. In cases of weighted systems, it is necessary to deal with
intensities rather than numbers of occurrence. Moreover, the latter
concept is considered as a special case of the former one. For the
authors, the triangles are among the simplest nontrivial motifs and
have a crucial role as one of the classic quantities of network
characterization in defining the clustering coefficient of a node
$i$. They propose a weighted clustering coefficient taking into
consideration the subgraph intensity, which is defined as the
geometric average of subgraph edge weights. In formula:

\begin{equation}
\label{CiO} \tilde{C}_{i,O} = \frac{2}{d_i(d_i-1)}\sum_{j,k\in V}
(\hat{w}_{ij}\hat{w}_{ik}\hat{w}_{jk})^{1/3}
\end{equation}

\noindent where  $\hat{w}_{ij} = w_{ij}/\max_{j\in V}(w_{ij})$ are
the edge weights normalized by the maximum weight in the network of
the edges linking $i$ to the other nodes of $V$.

Formula (\ref{CiO}) shows that triangles contribute to the creation
of $\tilde{C}_{i,O}$ according to the weights associated to their
three edges. More specifically, $\tilde{C}_{i,O}$ disregards the
strength of the local node and measures triangle weights only in
relation to the maximum edge weight.

Moreover, $\tilde{C}_{i,O}$ collapses to $C_i$ when, for each $i,
j\in V$, one has $w_{ij}=a_{ij}$, and is thus in the unweighted
case.

%%%%%%%%%%%%%%%%%%%%%%
\section{The generalized clustering coefficient}\label{model}

This section contains our proposal for a new definition of the
clustering coefficient of weighted networks. Based on a novel concept of triangles, this
definition includes the presence of real indirect connections among individuals. For our
purpose, we first provide and discuss the definition of the triangles, and then
we introduce the clustering coefficient.

\subsection{Generalized triples}
Here, we propose a generalization of the concept of triangle, and
rewrite accordingly the coefficient $C_i$ in (\ref{Ci}) for the case
of weighted networks.

\begin{definition}
\label{triangolo} Let us consider a weighted non-oriented graph
$G=(V,E)$ with vertices $V=\{1,\dots, N\}$, symmetric adjacent
matrix $\mathbf{A}=(a_{ij})_{i,j=1,\dots,N}$ and weight matrix
$\mathbf{W}=(w_{ij})_{i,j=1,\dots,N}$, with nonnegative weights.
Moreover, let us take $\alpha,\beta \in [0, \infty)$ and a function
$F:[0,+\infty)^2 \to [0,+\infty)$ which is not decreasing in its
arguments.

For each triple of distinct vertices $i, j, k \in V$, a subgraph
$t=(\{i,j,k\},E_T)$ is a generalized triangle (or, simply, a
triangle) around $i$ if one of the following conditions are
satisfied:
\begin{itemize}
\item[$T_1$] $a_{ij}=a_{ik}=a_{jk}=1$;
\item[$T_2$] $a_{ij}=a_{ik}=1$, $a_{jk}=0$ and $F(w_{ij},w_{ik}) \geq \alpha$;
\item[$T_3$] $a_{ij}=a_{jk}=1$, $a_{ik}=0$ and $F(w_{ij},w_{jk}) \geq \beta$.
\end{itemize}
\end{definition}

Herein we denote the elements of types $T_2$ and $T_3$ as {\it
triples} since they are not really triangles since they are not
contained in $G$. They can be seen as a generalization of triangles
by including the missing side, which is induced by conditions on the
weights of the two existing edges.

We denote the set of generalized triangles associated to case $T_h$
as $\mathcal{T}^{(i)}_h$, for $h=1, 2, 3$. By definition,
$\mathcal{T}_1^{(i)} \cap \mathcal{T}_2^{(i)}=\mathcal{T}_1^{(i)}
\cap \mathcal{T}_3^{(i)}=\mathcal{T}_2^{(i)} \cap
\mathcal{T}_3^{(i)}= \emptyset$. We denote the set collecting all
the triangles by $\mathcal{T}^{(i)}=\mathcal{T}_1^{(i)} \cup
\mathcal{T}_2^{(i)} \cup \mathcal{T}_3^{(i)}$.

\begin{figure}[t]
\begin{center}
\includegraphics[trim=0cm 1cm 0cm 0cm, clip = true, totalheight=0.15\textheight]{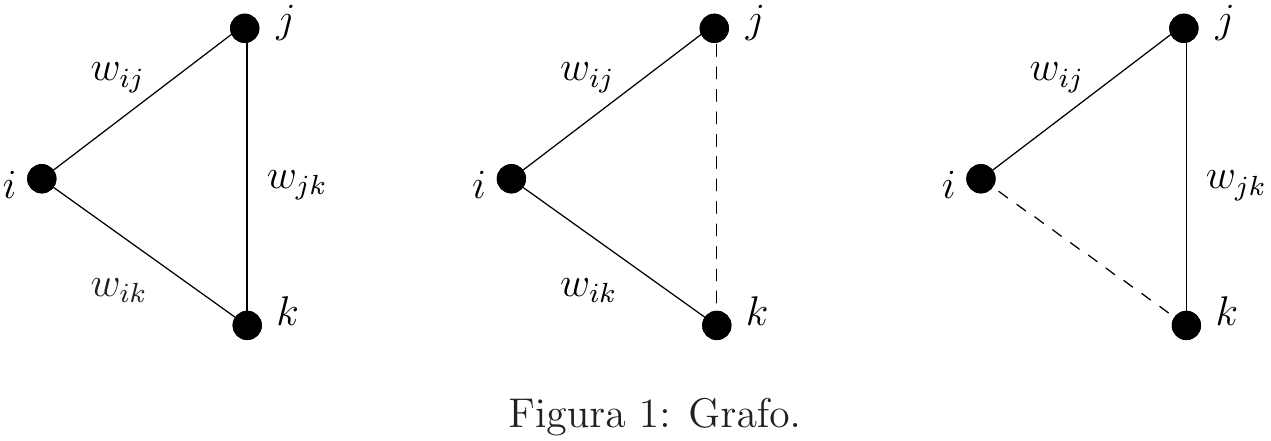}
\caption{Types of triangles: $T_1$ (left), $T_2$ (center), $T_3$ (right). }
\label{triangles}
\end{center}
\end{figure}

Figure~\ref{triangles} reports the three different type of
triangles, respectively $T_1$, $T_2$ and $T_3$. Clearly, in the case of
$T_1$, the concept of triangle given in Definition \ref{triangolo}
coincides with the standard one.

Note that with $N$ nodes, the maximum number of possible triangles
is $|\mathcal{T}_1|^* = \max |\mathcal{T}_1| = \binom{N}{3}$. This
is the case of a clique with $C_i = 1$, for each $i \in V$.

When considering the maximum number of candidates triangles for a
node $i$ to belong to $T_2$, it is $|\mathcal{T}_2^{(i)}|^* = \max
|\mathcal{T}_2^{(i)}| = \binom{d_i}{2}$. Then, in this case for the
node $i$ the number of triples is $|\mathcal{T}_2^{(i)}| =
|\mathcal{T}_2^{(i)}|^* - |\mathcal{T}_1^{(i)}|$.

Triples in $T_3$ for node $i$ are the paths of length $2$, % (i.e.
%triplets),
which can be computed by considering the square of the adjacency
matrix. Indeed, the number of different paths of length $2$ from $i$
to $k$ equals the entry $a_{ik}$ of $A^2$~\cite{Rosen}. For a given
row $i$ of $A^2$, the sum of the element (excluding the element
$a_{ii}$) equals the maximum potential number of triples of type
$T_3$.

Figure \ref{triangles} shows the types of triangles, without
emphasis on the conditions on the weights.

\subsection{Conceptualization of the generalized clustering coefficient}

Under Definition \ref{triangolo}, we can introduce a generalization
of the clustering coefficients presented in Formula (\ref{Ci}) for
weighted networks.

\begin{definition}
\label{ungc} Given a graph $G=(V,E)$ and a node $i \in V$, the
generalized unweighted clustering coefficient of $i$ is
\begin{equation}
\label{Cig} C_i^{(g)} = \frac{|\mathcal{T}^{(i)}|}{D_i}
\end{equation}
where $D_i = \frac{d_i(d_i-1)}{2} + |\{j\in V:
\Delta_{min}(i,j)=2\}|$, where $\Delta_{min}(i,j)$ is the minimum
distance between the nodes $i$ and $j$.
\end{definition}

The term unweighted in Definition \ref{ungc} points to the absence of
$w$'s in the coefficient in (\ref{Cig}). However, weights intervene
in the identification of the triangles, according to Definition
\ref{triangolo}. In particular, formula (\ref{Cig}) extends
(\ref{Ci}). As an example, notice that $C_i^{(g)}=C_i$ in the clique
case.

%%%%%%%%%%%
\subsection{Implications of the generalized clustering coefficient and equivalent graphs}
\label{implications}

The classical local clustering coefficient $C_i$ not only captures
the proportion of closed triples on all possible triples depending
on the degree of the ego/node $i$, but it also identifies its level
of cohesion. While the averaged clustering coefficient
$\overline{C}$ captures the whole level of network cohesion.

The proposed generalized clustering coefficient extends the same
setting also to the triples in $T_2$ and $T_3$, i.e. it is the
proportion of triangles of type $T_1$, $T_2$ and $T_3$ on all
possible triangles. This process depends not only on the degree of the ego but
also on the two thresholds $\alpha$ and $\beta$, which take into
account the strength profile around the ego, and thus have the possibility of creating
triangles $T_2$ and $T_3$.

The values of the generalized clustering coefficient are $0 \leq
C_i^{(g)} \leq 1$ as well as for the averaged measure $C^{(g)}$ and
differ from the usual measure because they depend on the thresholds
$\alpha$ and $\beta$.

Importantly, $C_i^{(g)}$ assumes unitary value not only in the
clique case, but also when any missing link is compensated by the
high weights of the other two links, i.e. when simultaneously
$\alpha < F(w_{ij},w_{ik}), \forall i, j, k$ and $\beta <
F(w_{ij},w_{jk}), \forall i, j, k$. This property of the generalized
clustering coefficient is very relevant, since it allows one to extend
the sense of community given by the classical clustering coefficient
to the case of indirect links being present, as seen in the
definition of triples $T_2$ and $T_3$.

The triples $T_2$ and $T_3$ can be described as follows (see
Figure~\ref{common}). The former describes a situation in which an
ego $i$ has a direct relationship with alters $j$ and $k$. One can
say that there exists a triangle among the three if the strength of
the connections of $i$ with the others is sufficiently high -- in
the sense described by function $F$. The idea is that the
cooperation and/or the common interests between $i$ and the alters
is so effective and fruitful that the presence of a direct link
between $j$ and $k$ is not required.

The latter case is associated to the presence of a strong link
between $i$ and $j$ and between $j$ and $k$, always in terms of the
entities of the weights -- in the sense described by function $F$.
In this peculiar situation, the node $j$ represents the intermediate
alter letting also the (indirect) collaboration between $i$ and $k$
be possible.

Finally, the thresholds have a double meaning. In fact, if we
consider a network in which interactions between alters could be facilitated,
an external decision-maker could implement policies aiming to define
the correspondent values of $\alpha$ and $\beta$ low. For example,
in the case of inter-organizational innovation networks, the presence of triangles
is positively related to the establishment of stable groups, as well as to the amount of produced
efforts, the straightening of transitive relationships and the
innovation capacity~\cite{CKL, FI1}.

On the other hand, if a decision-maker were to prevent interactions among alters,
the policies with correspondent values $\alpha$ and $\beta$ could be deemed sufficiently large.
Such an instance can be found in the prevention of community formation in criminal
organizations~\cite{FDCF, MB}.

\subsubsection{Equivalent graphs}

Triangles $T_1$, $T_2$ and $T_3$ also serve in deriving topological
information from the graph. In particular, assume that
$\alpha=\beta=0$, so that the number of  $T_2$ and $T_3$ around each
node does not depend on the specific selection of function $F$. In
this case, we know that $|\mathcal{T}_2^{(i)}|= \binom{d_i}{2}$, meaning
we are able to infer the degree of the node $i$ by the
knowledge of the number of triangles of type $T_2$ around it.
Conversely, $|\mathcal{T}_3^{(i)}|$ represents the number of
existing paths of length 2 having $i$ as one of the extreme nodes.
By collecting the number of the triangles $T_1$, $T_2$ and $T_3$ for
each node of the graph, we are able to identify a class of graphs.

Formally, consider a $3 \times N$ matrix collecting
$|\mathcal{T}_1^{(i)}|$, $|\mathcal{T}_2^{(i)}|$ and
$|\mathcal{T}_3^{(i)}|$, for each node $i \in V$. Denote by
$\mathcal{M}^{3,N}(\mathbb{N})$ the set of all the matrices with
dimension $3 \times N$ and filled by integer nonnegative numbers.

Thus, each matrix $\mathbf{M} \in \mathcal{M}^{3,N}(\mathbb{N})$
identifies a non-unique graph that has $N$ nodes and edges
described by $\mathbf{M}$. We refer to such a matrix as the
{\it triangles matrix}. In this sense, $\mathbf{M}$ can be viewed
as an equivalent class in the set of the graph with $N$ nodes, where
two graphs $G_1$ and $G_2$ are said to be equivalent when they share
the same matrix $\mathbf{M}$.

\begin{figure}[t]
    \centering
    \begin{minipage}{0.5\textwidth}
        \centering
        \includegraphics[scale = 0.9, trim = 0cm 0cm 0cm 0cm, clip]{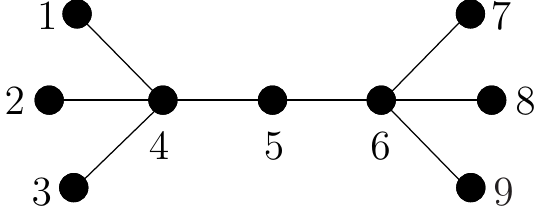}
        %\caption{a}
    \end{minipage}%
     %\ \hspace{5mm} \hspace{5mm}
    \begin{minipage}{0.5\textwidth}
        \centering
        \includegraphics[scale = 0.9, trim = 0cm 0cm 0cm 0cm, clip]{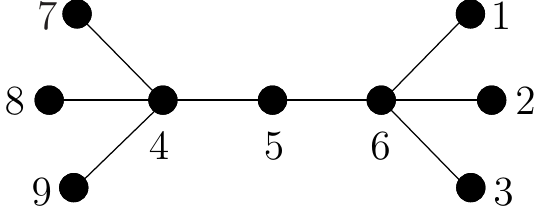}
        %\caption{b}
    \end{minipage}
    \caption{Two equivalent graphs, according to the equivalence defined through the triangles matrix.
    In this case, the triangles matrix $\mathbf{M}$ associated to the graphs is given in Table \ref{tab_trmat}.}
    \label{gclass}
\end{figure}

Figure \ref{gclass} and the matrix in (\ref{tab_trmat}) provide an
example of two equivalent classes, along with their common triangles
matrix $\mathbf{M}$. In particular, notice that matrix $\mathbf{M}$ is
the same for the two considered graphs, thus suggesting that the
equivalent class identified by $\mathbf{M}$ contains more than one
graph.

%
%\begin{equation}
%\label{tab_trmat}

\begin{table}[t]
%\begin{footnotesize}
%\begin{tiny}
\begin{center}
\begin{tabular}{|c|c|c|c|}
 \hline
  % after \\: \hline or \cline{col1-col2} \cline{col3-col4} ...
  node $i$ & $|\mathcal{T}_1^{(i)}|$ & $|\mathcal{T}_2^{(i)}|$ & $|\mathcal{T}_3^{(i)}|$
  \\ \hline
    1 & 0 & 0 & 3 \\
    2 & 0 & 0 & 3 \\
    3 & 0 & 0 & 3 \\
    4 & 0 & 5 & 1 \\
    5 & 0 & 1 & 6 \\
    6 & 0 & 5 & 1 \\
    7 & 0 & 0 & 3 \\
    8 & 0 & 0 & 3 \\
    9 & 0 & 0 & 3 \\
  \hline
\end{tabular}
\end{center}
\caption{Triangles matrix $\mathbf{M}$ associated with the graphs in
Figure \ref{gclass}.} \label{tab_trmat}
%\end{tiny}
%\end{footnotesize}
\end{table}

%%%%%%%%%%%%%%%%%%%%%%

\section{Applications}

Herein we considered the analysis of the generalized clustering
coefficient on two empirical networks: the network among the $500$ busiest US
commercial airports~\cite{CPV} and the nervous system of the
nematode Caenorhabditis elegans~\cite{WS, WST}. The data processing, the
network analysis and all simulations were conducted using the
software {\it R}~\cite{RCT} with the {\it igraph} package~\cite{CN}.  The datasets were obtained from the $R$
packege {\it tnet}, authored by Tore Opsahl~\cite{opsahl}.
Code in the $R$ programming language is available upon request.

For the sake of readability we report in Table~\ref{tab_2} the notations used hereafter.

\begin{table}[t]
\begin{footnotesize}
%\begin{tiny}
\begin{center}
\begin{tabular}{|c|l|}
\hline
Symbol & Meaning\\
\hline\hline
$t_i$                               & Triangles around node $i$.\\
$T_h$                           & Triples of type $h=1, 2, 3$. \\
$\mathcal{T}^{(i)}_h$         & Set of triples of node $i$ associated to case $T_h$, for $h=1, 2, 3$.\\
$|\mathcal{T}^{(i)}_h|$         & Cardinality of the set $\mathcal{T}^{(i)}_h$\\
$F_i$                           & Function of type $i= 1, 2, 3, 4.$\\
$\alpha$                        & Threshold for triples $T_2$\\
$\beta$                         & Threshold for triples $T_3$\\
$C_i$                           & Local clustering coefficient\\
$\overline{C}$                  & Averaged clustering coefficient\\
$C_G$                           & Global clustering coefficient\\
$C_i^{(g)}$                     & Generalized clustering coefficient\\
\hline
\hline
\end{tabular}
\end{center}
\caption{Table of notation.}
\label{tab_2}
%\end{tiny}
\end{footnotesize}
\end{table}

%%%%% Inizio US airport %%%%%%%
\subsection{General settings}

In the empirical experiments, we consider four cases of function $F$:

\begin{itemize}
\item[$F_1$] sum of the weights is greater than the correspondent coefficient: $w_{ij} + w_{ik} \geq \alpha$ and $w_{ij} + w_{jk} \geq
\beta$;
\item[$F_2$] average of the weights is greater than the correspondent coefficient: $(w_{ij} + w_{ik})/2 \geq \alpha$ and $(w_{ij} + w_{jk})/2 \geq
\beta$;
\item[$F_3$] minimum of the weights is greater than the correspondent coefficient: $\min\{w_{ij},w_{ik}\} \geq \alpha$ and $\min\{w_{ij},w_{jk}\} \geq
\beta$;
\item[$F_4$] maximum of the weights is greater than the correspondent coefficient: $\max\{w_{ij}, w_{ik}\} \geq \alpha$ and $\max\{w_{ij}, w_{jk}\} \geq
\beta$.
\end{itemize}

The selection of the specific function $F$ -- to be implemented
among $F_1, \ldots, F_4$ defined above -- provides further insights
into the interpretation of the triples of type $T_2$ and $T_3$.
Indeed, once $\alpha$ and $\beta$ are kept fixed, then $F_1$ and
$F_2$ state that both weights of the considered edges should be
taken into account in an identical way by considering their mere
aggregation in the former case or their mean in the latter one. When
considering functions $F_3$ and $F_4$, only one of the weights is
relevant for the measurement of the strength of the connections --
the minimum weight and the maximum one, respectively. Naturally, the
former case is more restrictive than the latter one, since it
implicitly assumes that both weights should be greater than $\alpha$
or $\beta$ for having a triples of type $T_2$ or $T_3$.

Social sciences suggest other functions $F$'s to be considered
in Definition (\ref{triangolo}) to capture certain
peculiarities of the system under observation. Notice also that
$|\mathcal{T}_2^{(i)}|$ and $|\mathcal{T}_3^{(i)}|$ are not
increasing functions of $\alpha$ and $\beta$, respectively, as
Definition (\ref{triangolo}) implies.

For the simulations, the value of $\alpha$ and $\beta$ are $\alpha, \beta = \{ 0, 250000, 500000,
750000, 1000000,$ $1250000, 1500000, 1750000, 2000000, 2225000\}$
for the US airports network and  $\alpha, \beta =
\{ 0, 5, 10, 15, 20, 25, 30, 35, 40, 45, 50, 55, 60, 65, 70\}$ for
the C.elegans network. The max values were chosen on the ground that
function $F_1$ could possibly be true also when considering arcs
with the higher weights. As such, $10$ runs were implemented for
each considered value. Thus, we performed $100$ computations for the US airports network
and $150$ computations for the C.elegans network.

According to Definition (\ref{triangolo}), $\mathcal{T}_2^{(i)}$ and
$\mathcal{T}_3^{(i)}$, i.e. the triangles for every node in
a network, can be computed considering $\alpha = 0$ and $\beta = 0$.
Concerning the sets $\mathcal{T}_1^{(i)}$,
such triangles can be easily computed by a built-in function in
$igraph$.

%%%%% Inizio US airport %%%%%%%
\subsection{Analysis of the US commercial airports network}

The US commercial airports network has $n = 500$ nodes denoting
airports and $m = 2980$ edges representing flight connections. In
this network, weights are the number of seats available on that
connections in 2010. The network has both small-world and scale-free
organization with $\gamma \simeq 1.8$~\cite{BBPV2}.

\begin{figure}[!htb]
    \centering
    \begin{minipage}{0.5\textwidth}
        \centering
        \includegraphics[scale = 0.3, trim = 0cm 0cm 0cm 0cm, clip]{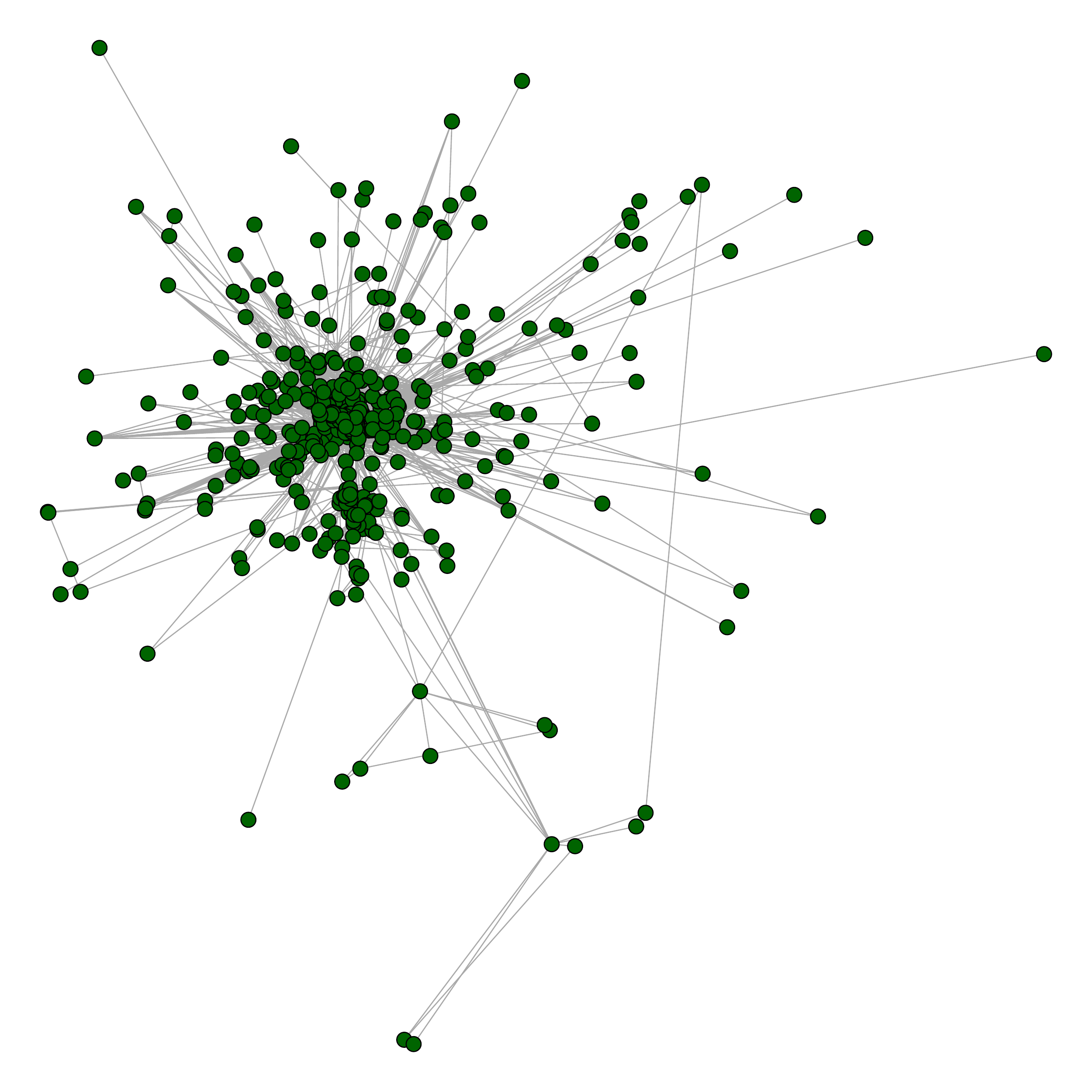}
        %\caption{a}
    \end{minipage}%
     %\ \hspace{5mm} \hspace{5mm}
    \begin{minipage}{0.5\textwidth}
        \centering
        \includegraphics[scale = 0.3, trim = 0cm 0cm 0cm 0cm, clip]{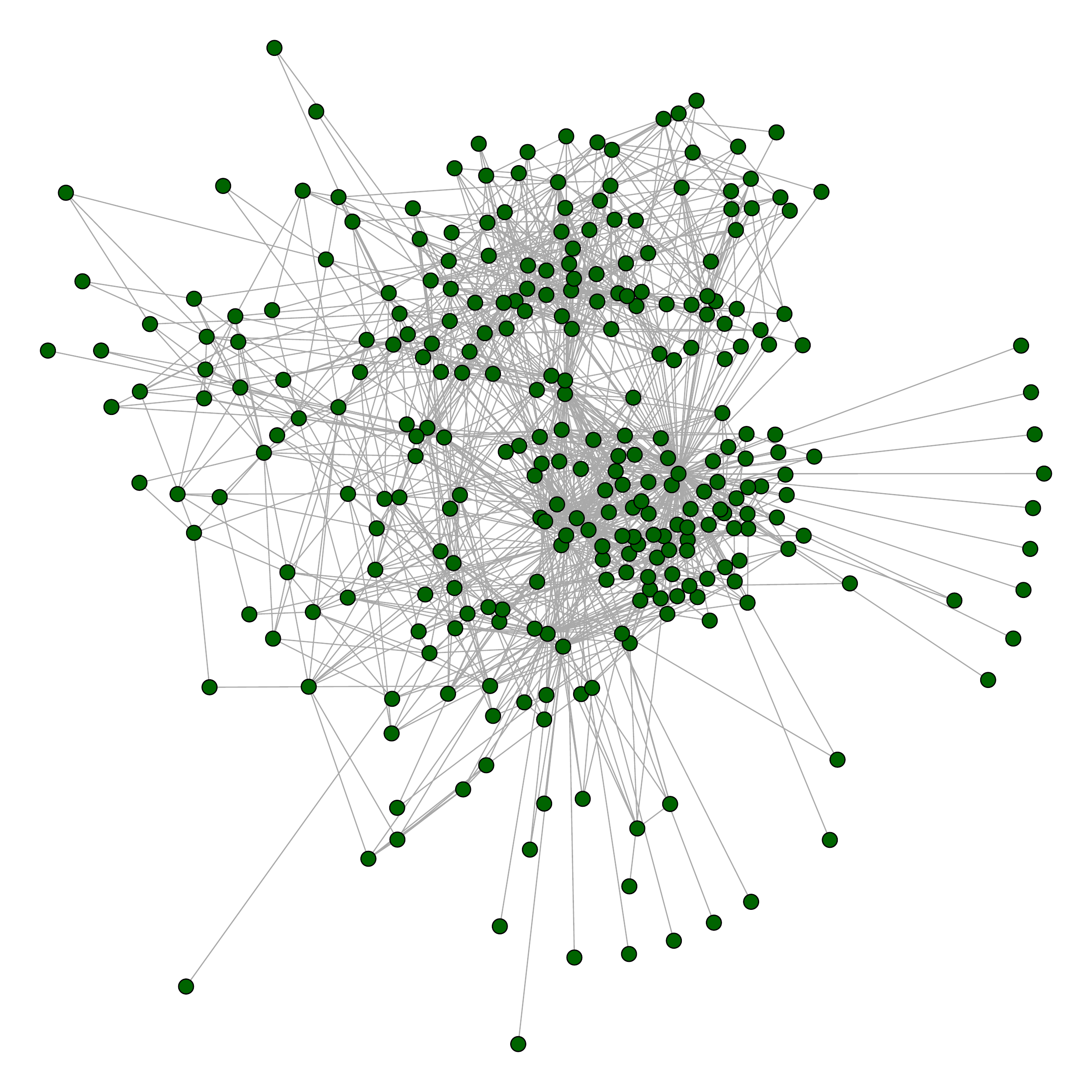}
        %\caption{b}
    \end{minipage}
    \caption{Network visualization for the US airports (left) and C.elegans (right).}
    \label{visualization}
\end{figure}

In Figure~\ref{visualization} (left) we show the network visualization,
while Table~\ref{tab_1} reports some basic measures: the
density $\delta$, the averaged clustering coefficient
$\overline{C}$, the global clustering coefficient $C_G$ and the
minimum, maximum and average degree, weight and strength.

\begin{figure}[!htb]
    \centering
    \begin{minipage}{0.5\textwidth}
        \centering
        \includegraphics[scale = 0.35, trim = 3cm 0cm 3cm 0cm, clip]{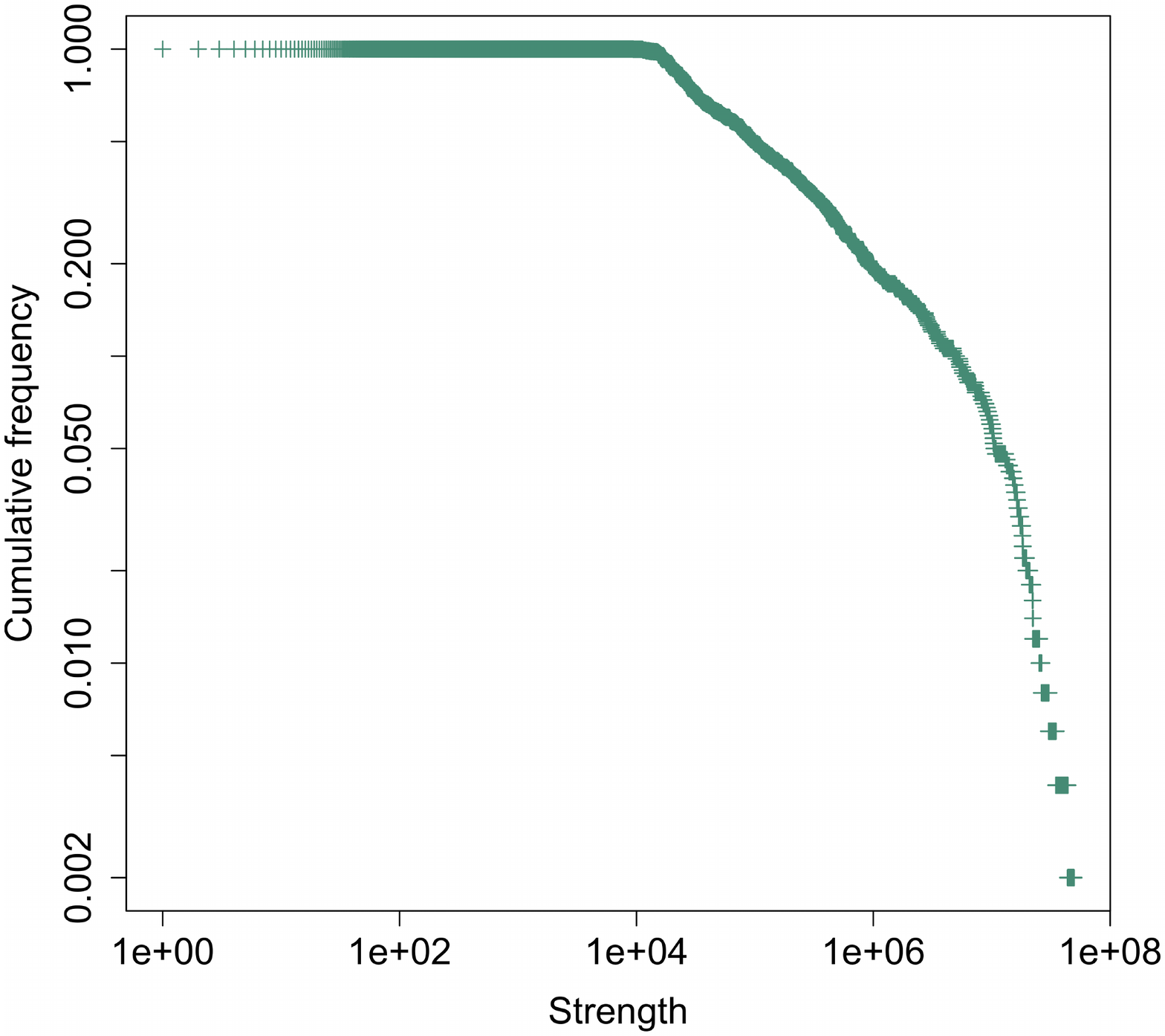}
        %\caption{a}
    \end{minipage}%
     %\ \hspace{5mm} \hspace{5mm}
    \begin{minipage}{0.5\textwidth}
        \centering
        \includegraphics[scale = 0.35, trim = 3cm 0cm 3cm 0cm, clip]{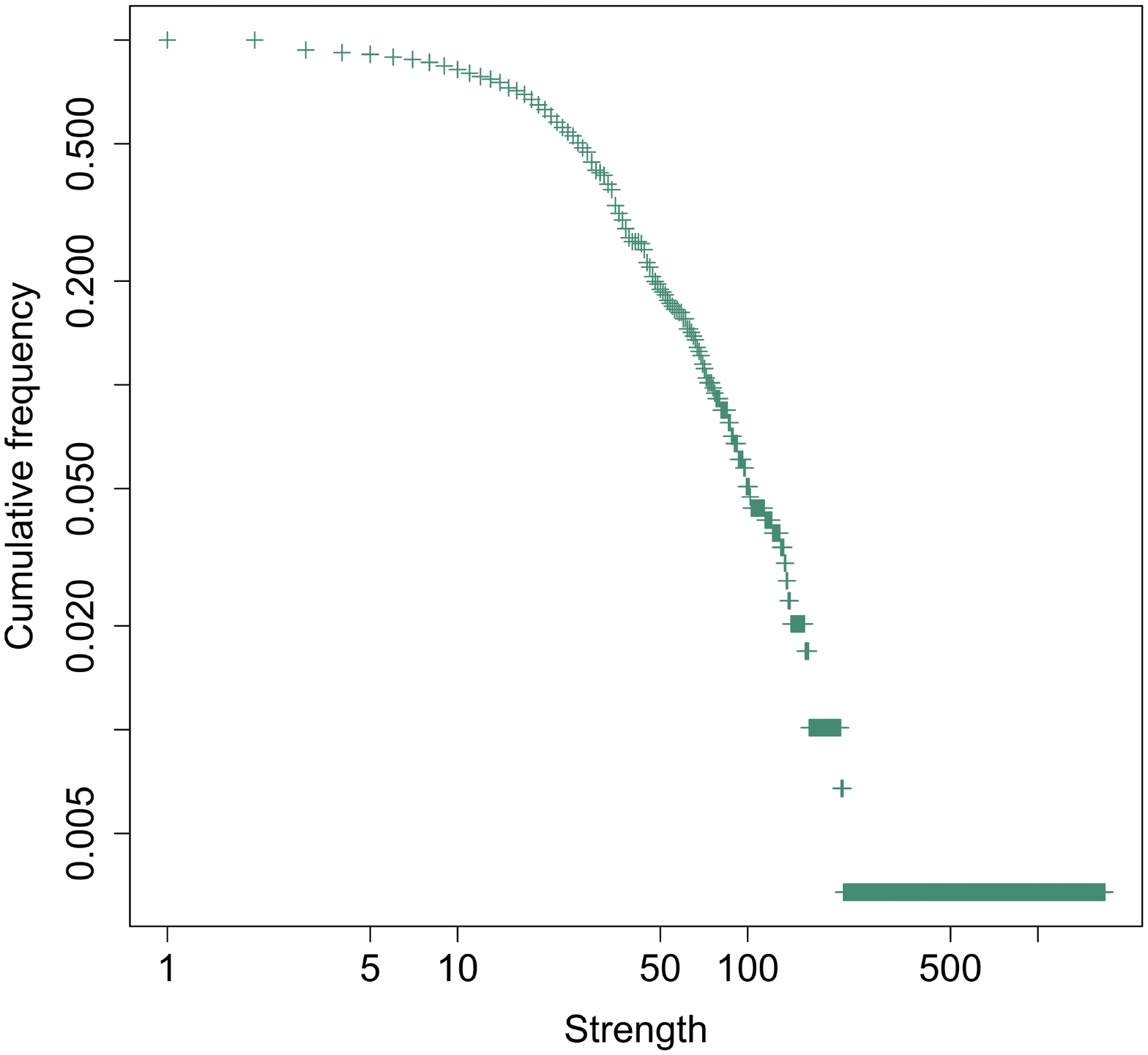}
        %\caption{b}
    \end{minipage}
    \caption{Strength distributions for the US airports (left) and C.elegans (right) networks.}
    \label{strength}
\end{figure}

\begin{figure}[!htb]
    \centering
    \begin{minipage}{0.5\textwidth}
        \centering
        \includegraphics[scale = 0.3, trim = 0cm 0cm 0cm 0cm, clip]{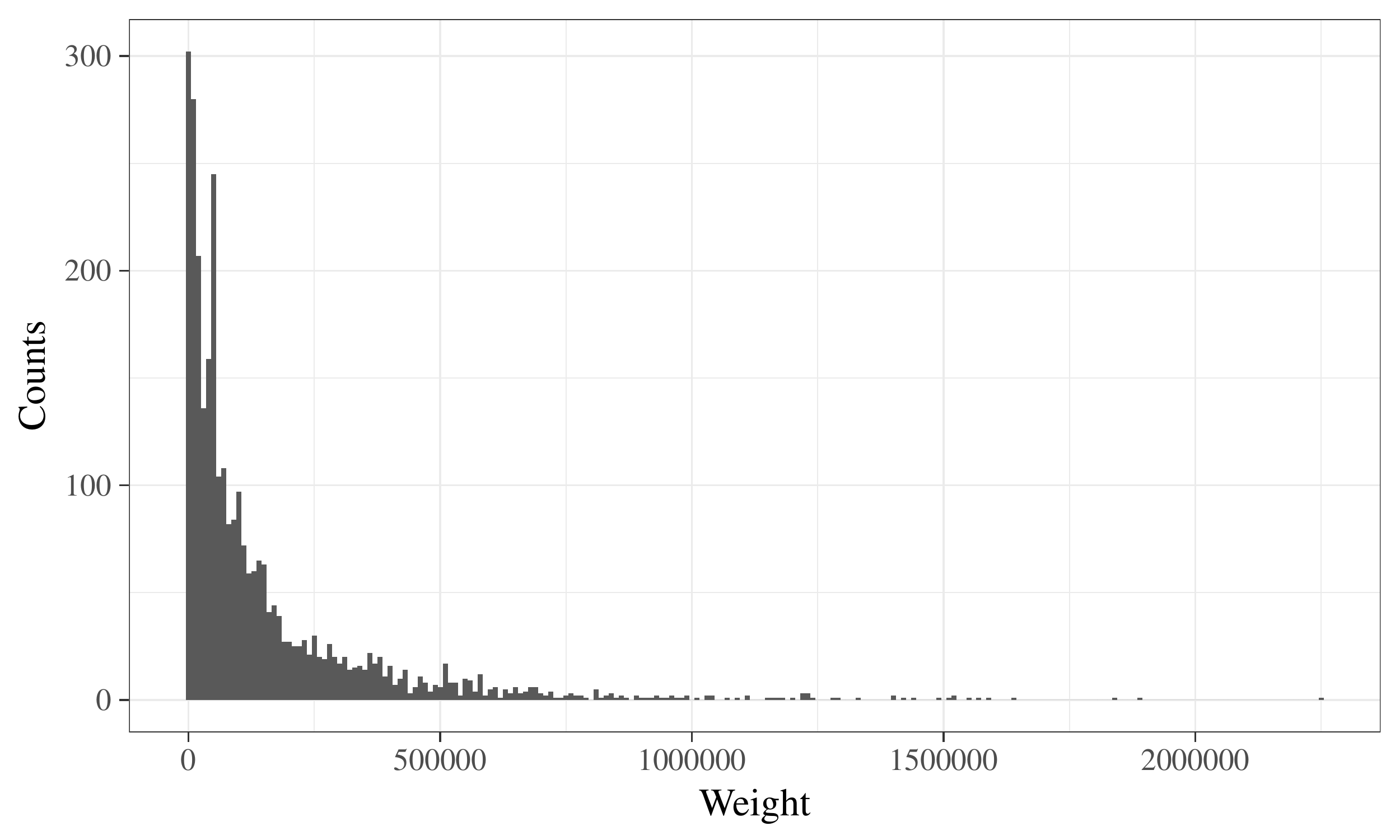}
        %\caption{a}
    \end{minipage}%
     %\ \hspace{5mm} \hspace{5mm}
    \begin{minipage}{0.5\textwidth}
        \centering
        \includegraphics[scale = 0.3, trim = 0cm 0cm 0cm 0cm, clip]{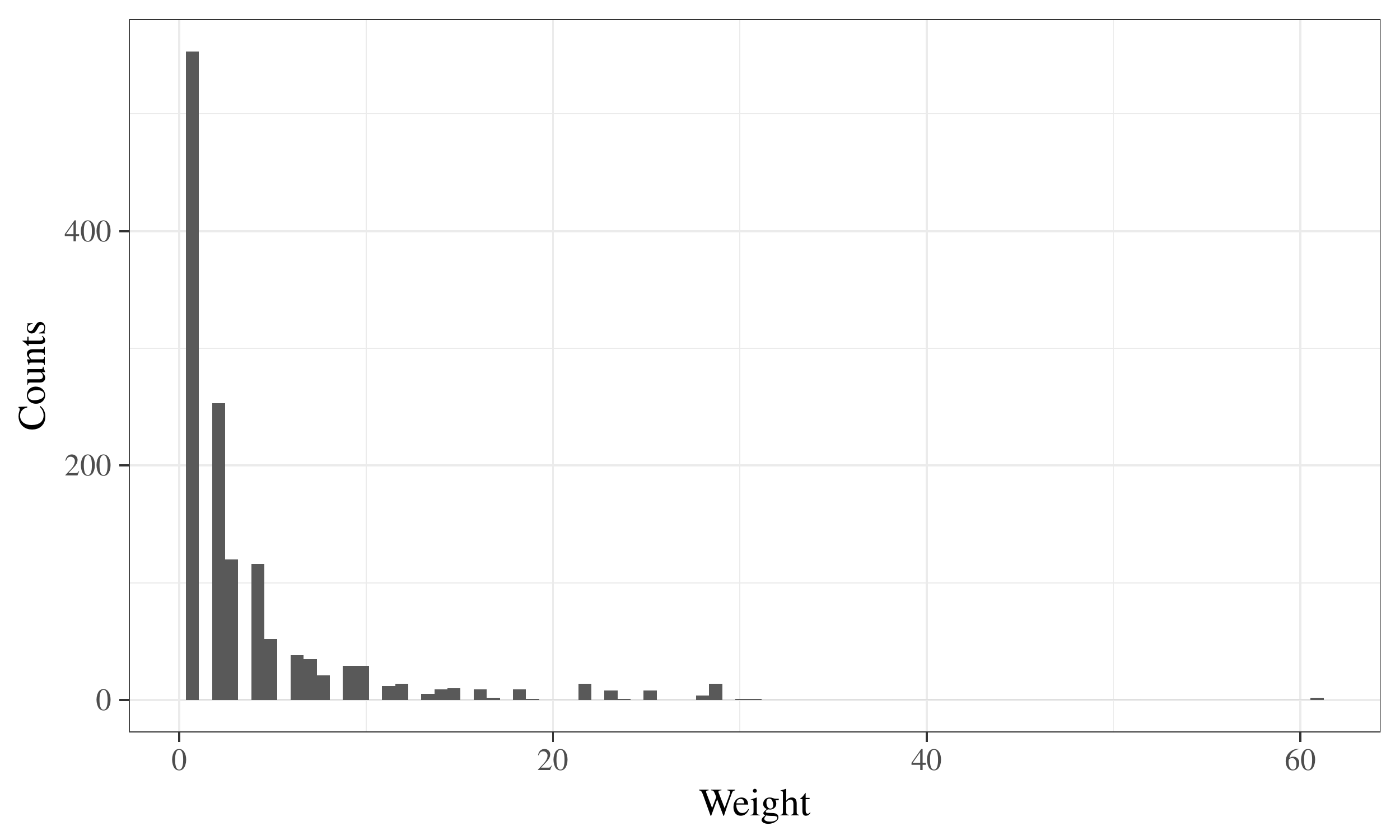}
        %\caption{b}
    \end{minipage}
    \caption{Histogram displaying weights for US airports (left) and C.elegans (right) networks.}
    \label{weight_hist}
\end{figure}

In Figure~\ref{strength} (left) we report the strength distribution for
this network, with the strength $s_i$ as the sum of the
weights of the links incident on $i$, while Figure~\ref{weight_hist} (left)
uses a histogram to display the weights.

\begin{table}[t]
\begin{footnotesize}
%\begin{tiny}
\begin{center}
\begin{tabular}{|l|c|c|c|c|c|c|c|c|c|c|c|c|}
\hline
Network & $\delta$ & $\overline{C}$ & $C_G$ & $k_{min}$ &  $k_{max}$ & $\overline{d}$\\
\hline\hline
US airports & 0.0239 & 0.617 & 0.351 & 1 & 145 & 11.92\\
C.elegans & 0.0314 & 0.228 & 0.121 & 1 & 134 & 9.26 \\
\hline
Network & $w_{min}$ & $w_{max}$ & $\overline{w}$ &$s_{min}$ & $s_{max}$ & $\overline{s}$\\
\hline\hline
US airports & 9 & 2253992 & 152320.19 & 9416 & 49316361 & 1815656.66\\
C.elegans & 1 &       61     & 4.198 &   1 & 1700 & 38.86\\
\hline
\end{tabular}
\end{center}
\caption{Basic measures for the networks under analysis.}
\label{tab_1}
%\end{tiny}
\end{footnotesize}
\end{table}

\begin{figure}[!htb]
    \centering
    \begin{minipage}{0.5\textwidth}
        \centering
        \includegraphics[scale = 0.3, trim = 0cm 0cm 0cm 0cm, clip]{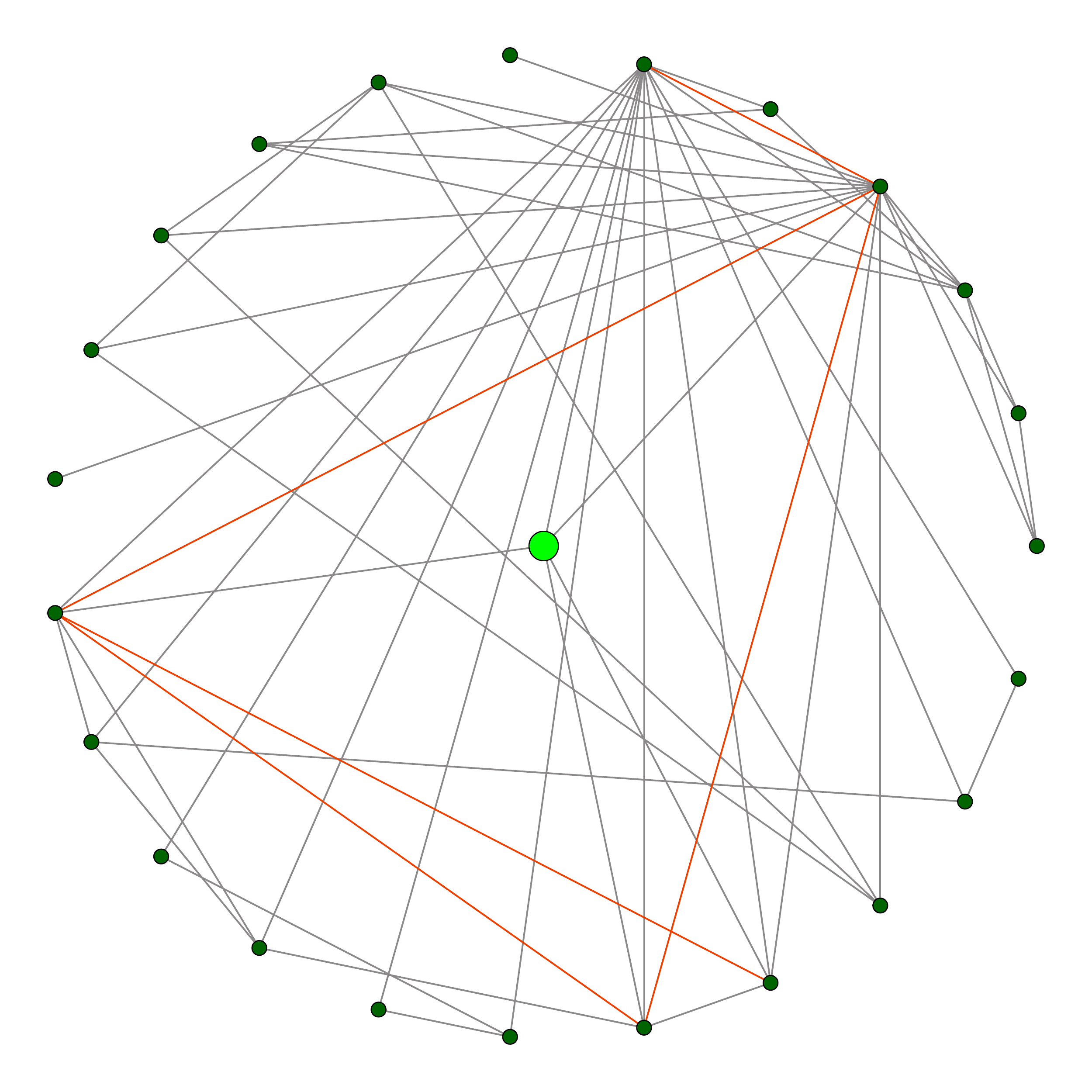}
        %\caption{a}
    \end{minipage}%
     %\ \hspace{5mm} \hspace{5mm}
    \begin{minipage}{0.5\textwidth}
        \centering
        \includegraphics[scale = 0.3, trim = 0cm 0cm 0cm 0cm, clip]{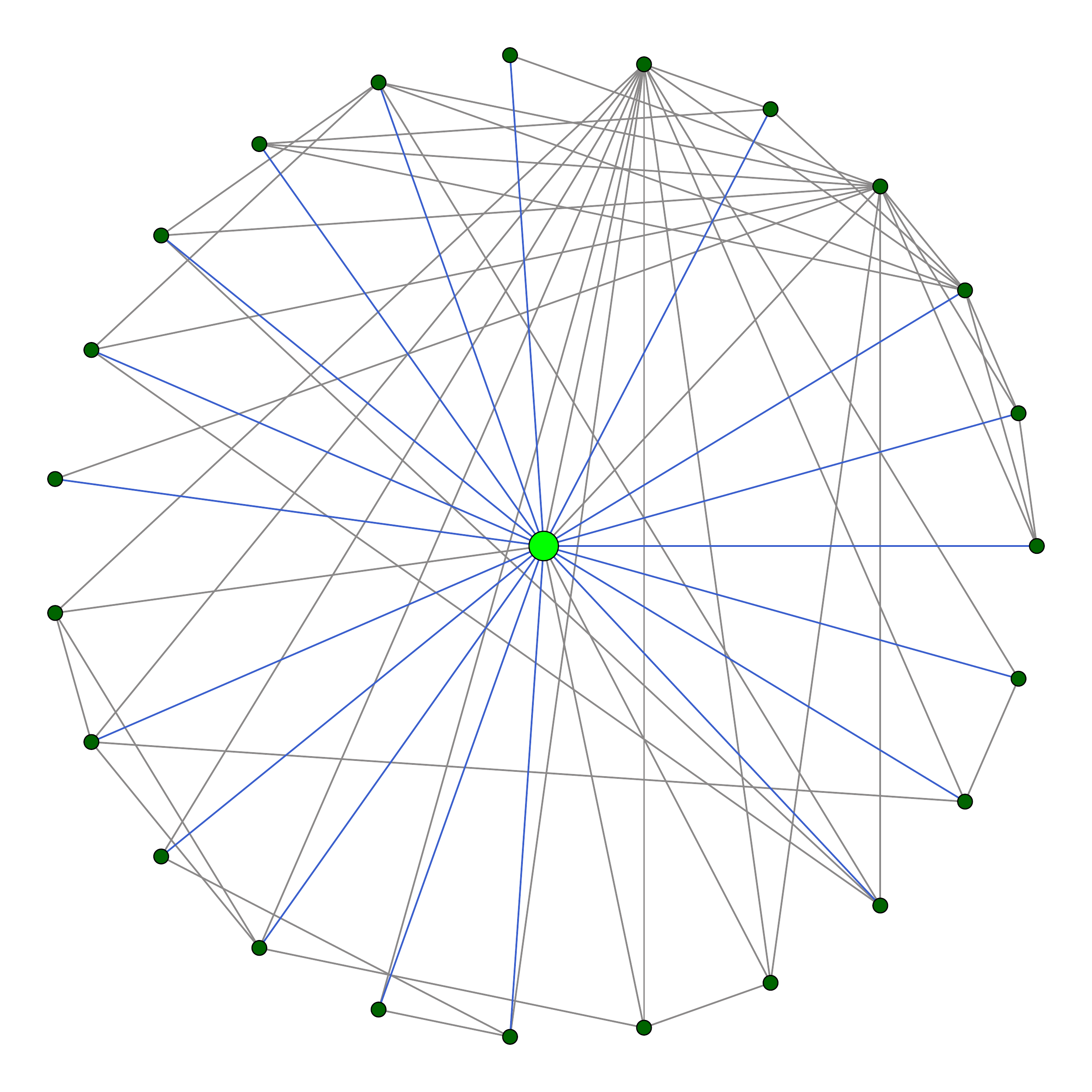}
        %\caption{b}
    \end{minipage}
    \caption{$2-$step ego network of node $n = 488$ of the US airports
    network and triples $\mathcal{T}_2^{(488)}$ (left) and $\mathcal{T}_3^{(488)}$ (right).}
    \label{triangles488}
\end{figure}

Functioning as an example, Figure~\ref{triangles488} shows the arcs
composing the triples in $\mathcal{T}_2^{(488)}$ and in
$\mathcal{T}_3^{(488)}$ for the neighborhood of order $2$ of node $n
= 488$, i.e. its 2-step ego network. Such a node has a degree
$d_{488} = 5$, a second order neighborhood of cardinality $18$ and a
local clustering coefficient $C_{488} = 0.5$, because $5$ triangles are closed 
out of a theoretical $10$.

Thus, $|\mathcal{T}_2^{(488)}| = 5$
while triangles in $\mathcal{T}_3^{(488)}$ are computed obtaining
$|\mathcal{T}_3^{(488)}| = 22$. Note that the blue arcs in the right
panel of Figure~\ref{triangles488} are $18 (< 22)$ because some arcs
can be mentioned twice in the set, since arc $(i, k)$ can derive
from $i \rightarrow j \rightarrow k$ as well as from $i \rightarrow
l \rightarrow k$.

The generalized clustering coefficient has value  $C_{488}^{(g)} =
0.0632$, which is much lower than $C_{488}$ since the proportion of closed
triangles when $\alpha = 0$ and $\beta = 0$ is
smaller than the basic setting.

\begin{figure}[htbp]
\begin{center}
\includegraphics[scale=.5]{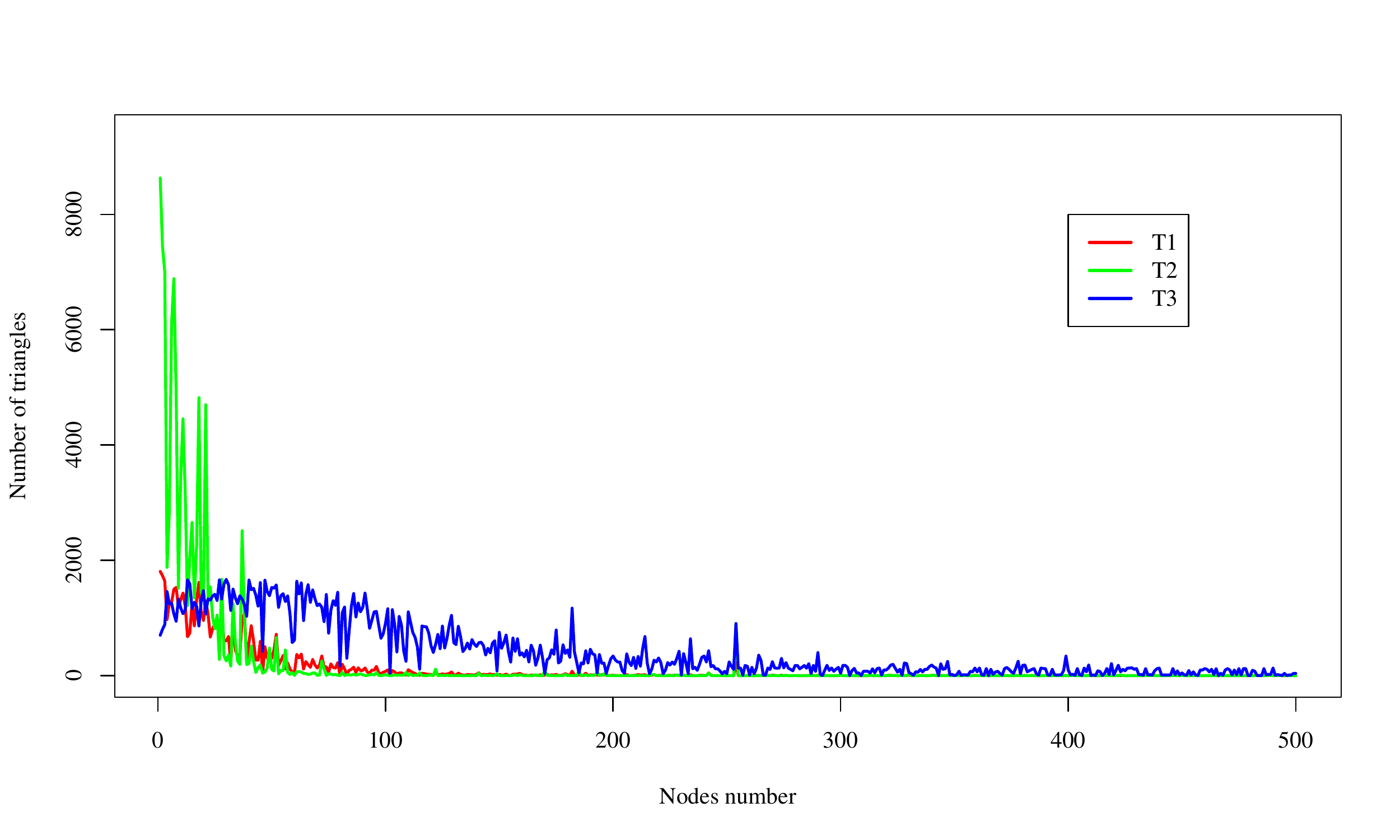}
\caption{US airports. Comparison between the number of triangles
$T_1$, the number of potential triples $T_2$ and the number
of potential triples $T_3$.} \label{usadeltat}
\end{center}
\end{figure}

\begin{figure}[htbp]
\begin{center}
\includegraphics[scale=.5]{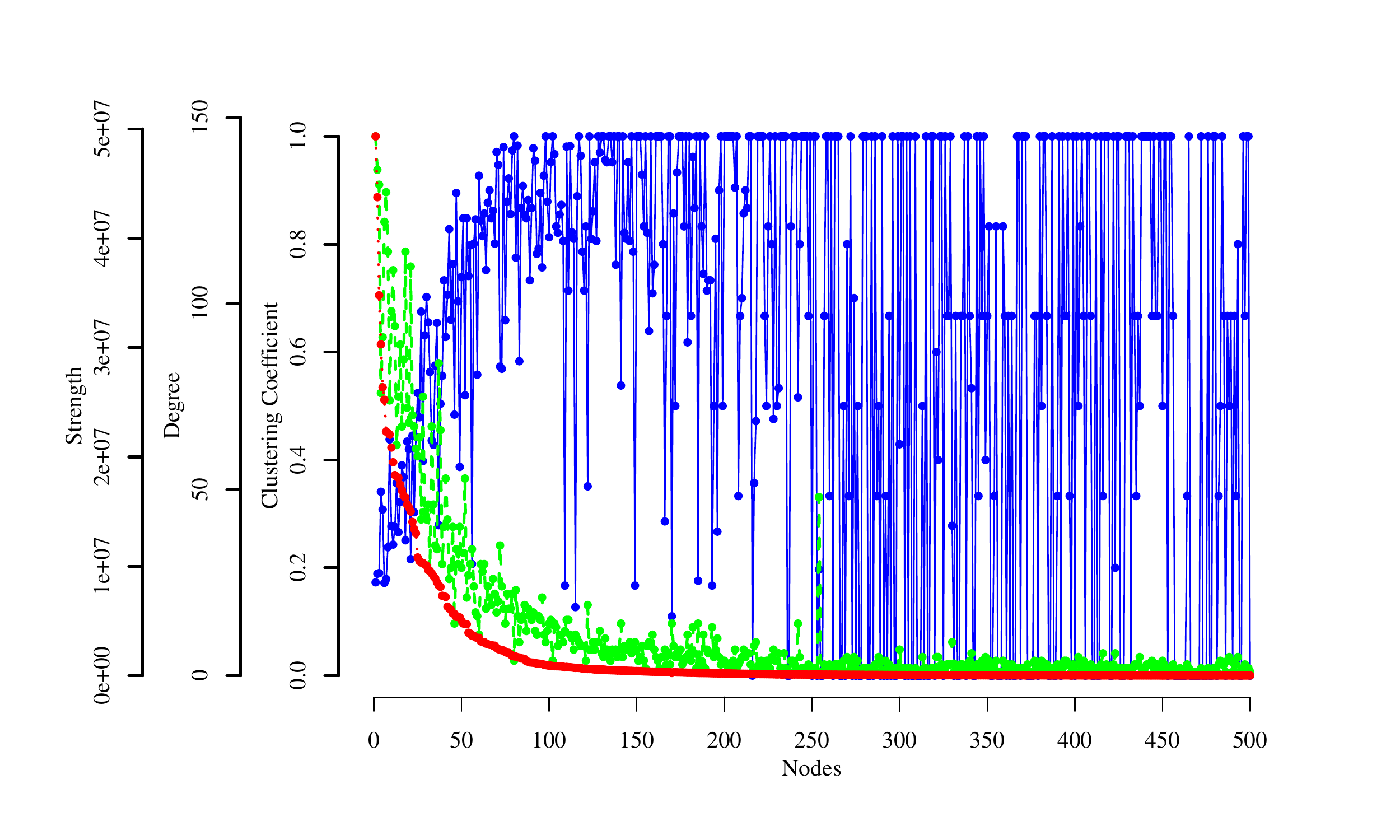}
\caption{US airports. Comparison between local clustering coefficient (blue points), degree (red points) and strength (green points).}
\label{usaTDS}
\end{center}
\end{figure}

Figure~\ref{usadeltat} for the US airports network reports three curves for
each node: the total number of triangles $|\mathcal{T}_1^{(i)}|$,
the number of potential triples of type
$|\mathcal{T}_2^{(i)}|$ and the number of potential triples
of type $|\mathcal{T}_3^{(i)}|$. Figure~\ref{usaTDS} compares
the degree $d_i$ and the local clustering
coefficient $C_i$ for each node $i$. Note that nodes in the US airports network are
enumerated in non-increasing order of their degree and the nodes with indices until
$i \simeq 100$ have values of degree and clustering
coefficient, which allow for a large number of triples $T_2$
and a significant number of triples $T_3$. Then, when the
degree decreases and the local clustering coefficient increases, the
local neighborhoods preclude the formation of triangles.

\begin{figure}[!htb]
    \centering
    \begin{minipage}{0.5\textwidth}
        \centering
        \includegraphics[scale = 0.4, trim = 5cm 0cm 5cm 0cm, clip]{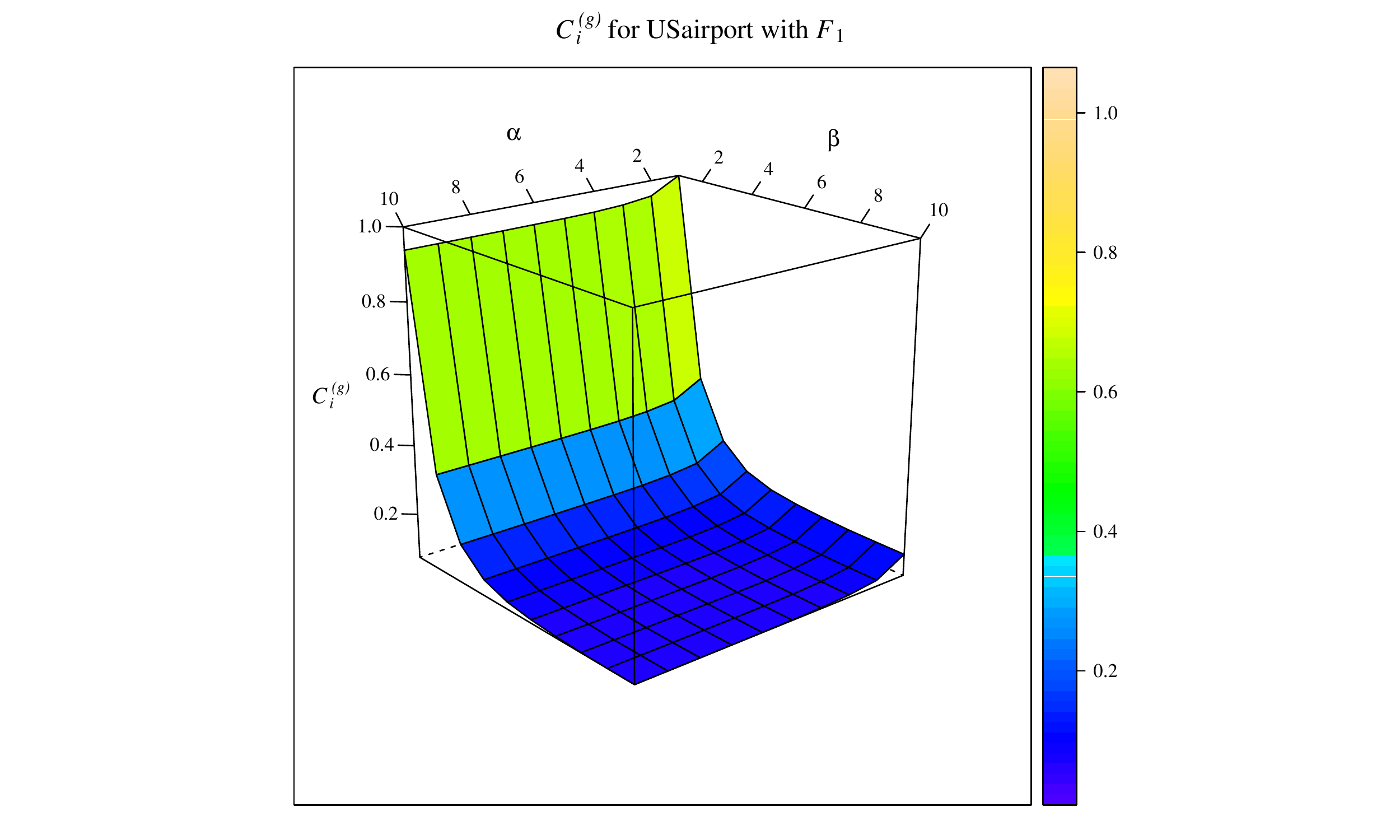}
        %\caption{a}
    \end{minipage}%
     %\ \hspace{5mm} \hspace{5mm}
    \begin{minipage}{0.5\textwidth}
        \centering
        \includegraphics[scale = 0.4, trim = 5cm 0cm 5cm 0cm, clip]{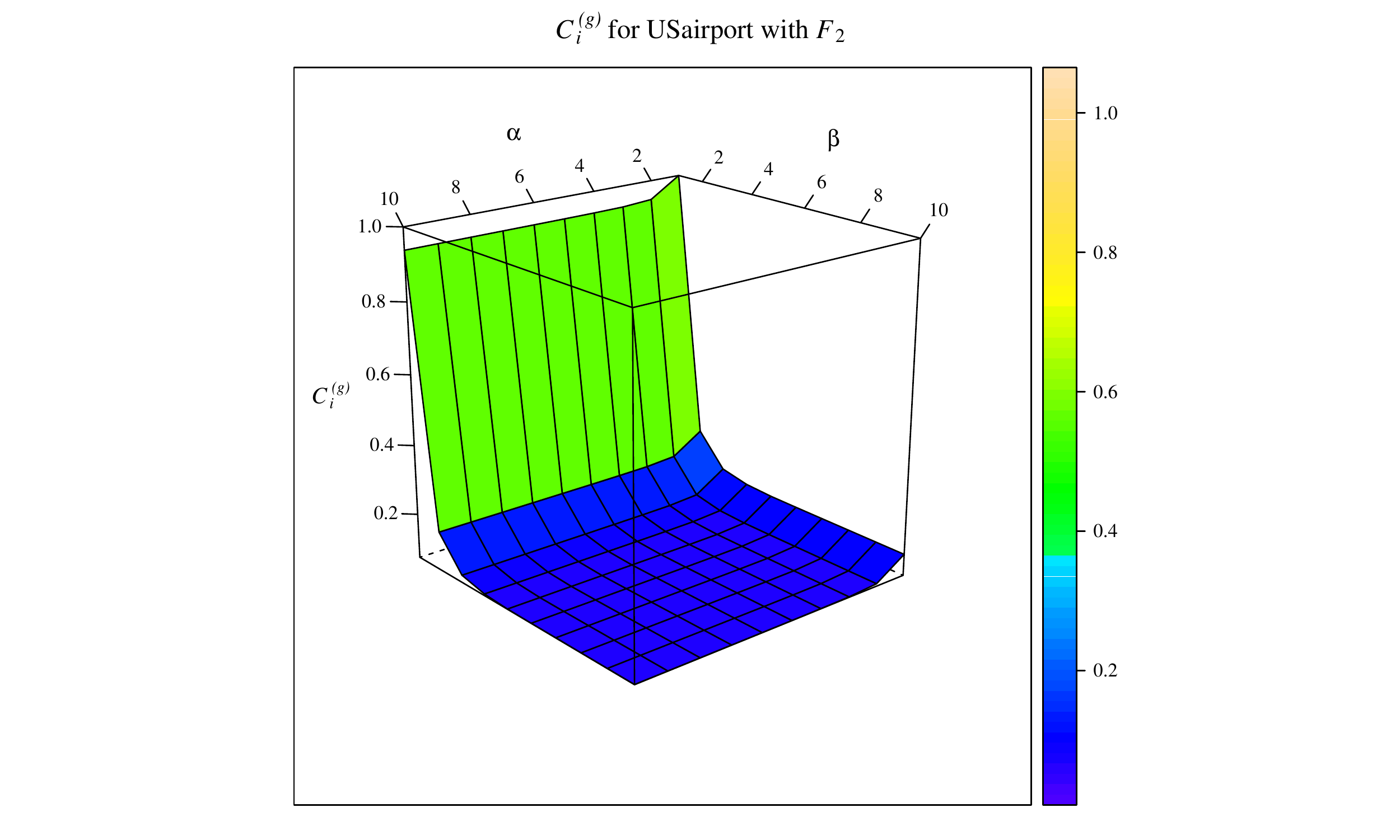}
        %\caption{b}
    \end{minipage}
        \begin{minipage}{0.5\textwidth}
        \centering
        \includegraphics[scale = 0.4, trim = 5cm 0cm 5cm 0cm, clip]{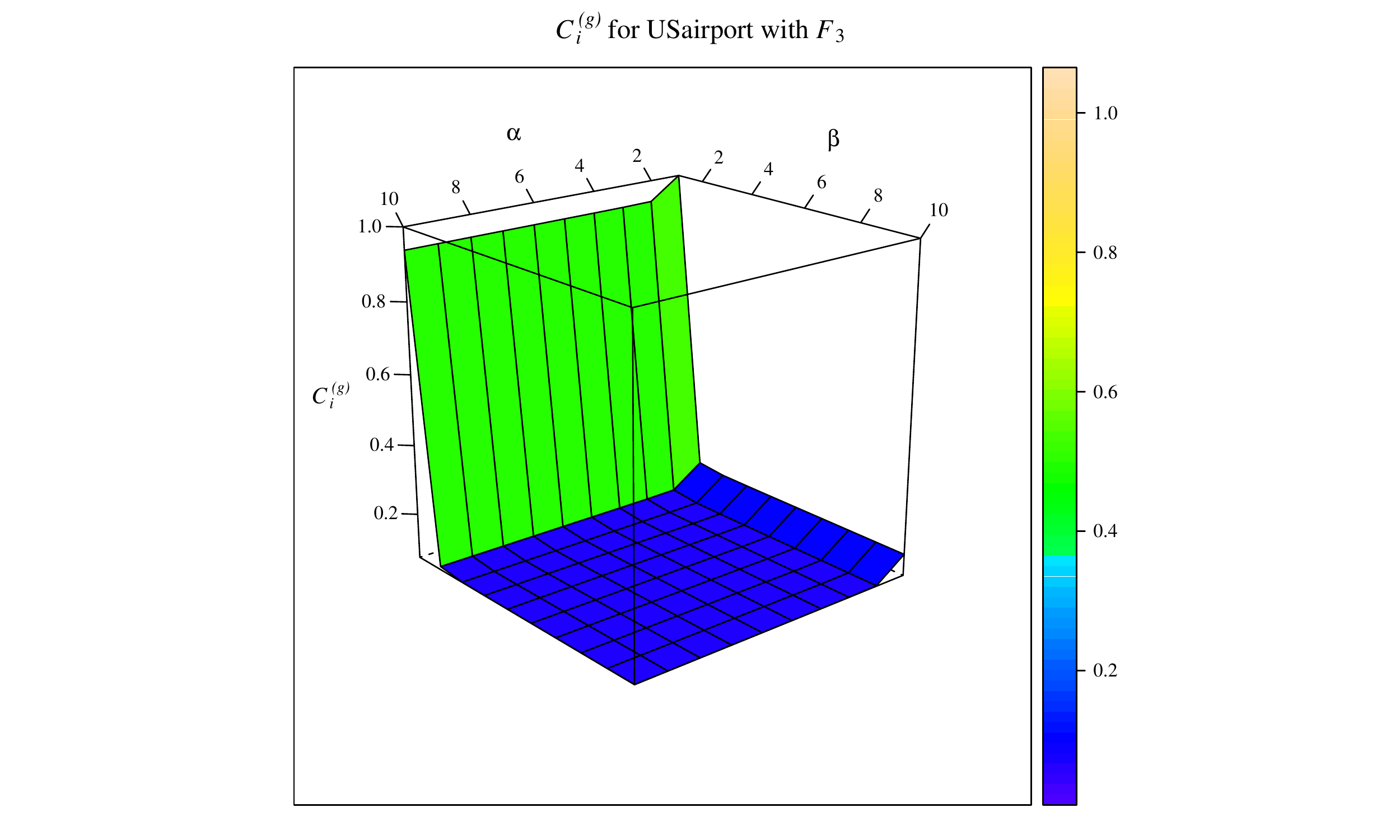}
        %\caption{c}
    \end{minipage}%
     %\ \hspace{5mm} \hspace{5mm}
    \begin{minipage}{0.5\textwidth}
        \centering
        \includegraphics[scale = 0.4, trim = 5cm 0cm 5cm 0cm, clip]{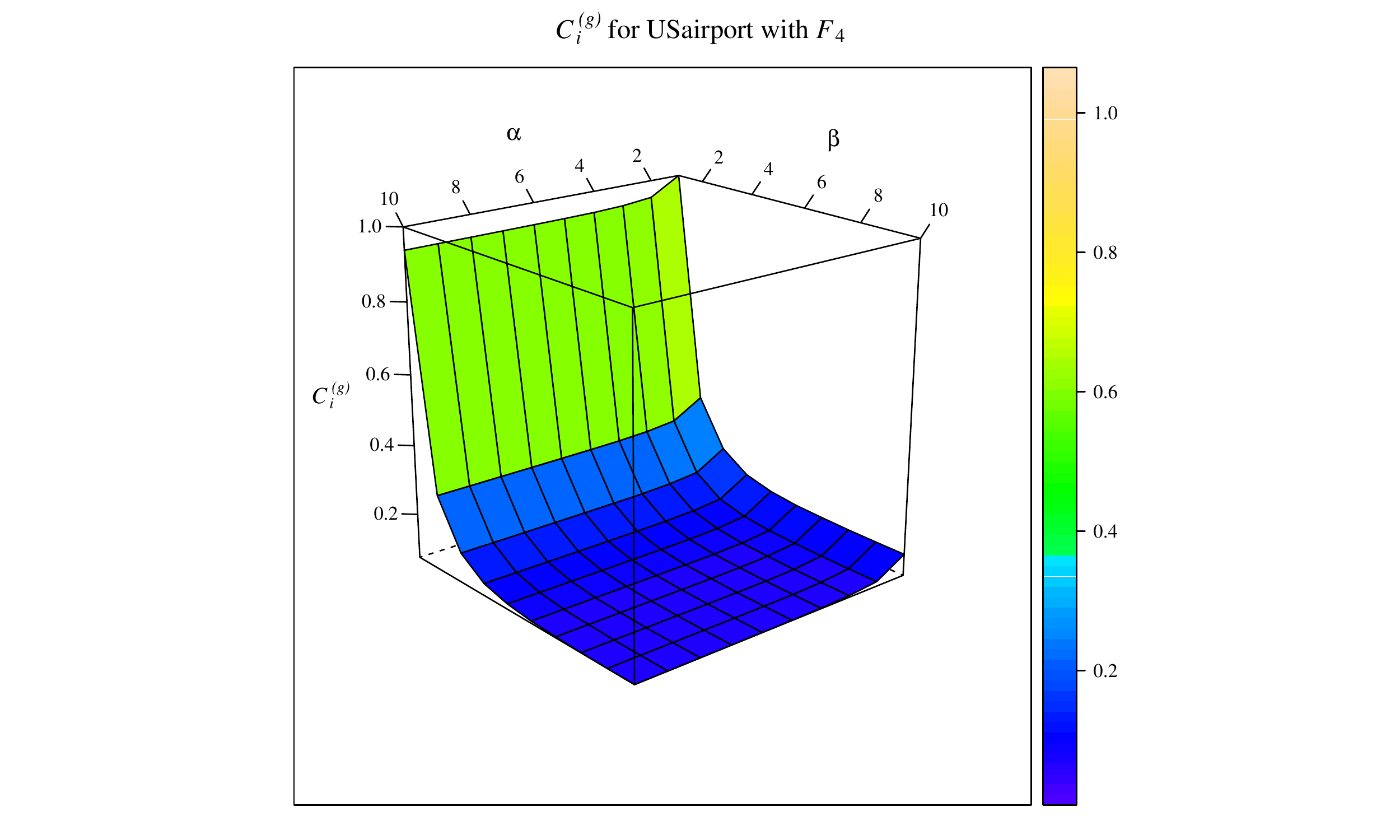}
        %\caption{d}
    \end{minipage}
    \caption{US airports: Average values of $C_i^{(g)}$ for cases $F_1$ (upper left), $F_2$ (upper right), $F_3$ (lower left) and $F_4$ (lower right).}
    \label{usaCg}
\end{figure}

Figure~\ref{usaCg} shows the averaged values of the generalized
clustering coefficient $C_i^{(g)}$ for the US airports network when
considering the four different functions $F_1, F_2, F_3$ and $F_4$.
In each figure, the values are presented for every combination of
$\alpha$ and $\beta$ while the horizontal axis reports the values of
$C_i^{(g)}$ as averaged over every node in the network.
As expected, higher values of $C_i^{(g)}$ are obtained for lower
values of $\alpha$ and $\beta$ and, globally, we have a non-increasing
trend with a higher slope for functions $F_2$ and $F_3$
since the average and the min functions smooth the values, thus
indicating that the functions are true only for small values of weights.
Regarding $F_1$ and $F_4$, they are more prone to being true for higher
values of arc weight, meaning the slope declines at slower rate.

A common behavior for all four cases is that the magnitude of
$C_i^{(g)}$ is more dependent on triples $T_2$ than those
in $T_3$. This is due to the tendency of high-degree nodes to have
a higher strength. Therefore, the functions are more prone to being
true for triples $T_2$ than for triples in $T_3$
since the adjacent links could possibly lie in a low-degree node
with a low value of strength.

\begin{figure}[!htb]
    \centering
    \begin{minipage}{0.5\textwidth}
        \centering
        \includegraphics[scale = 0.3, trim = 0cm 0cm 0cm 0cm, clip]{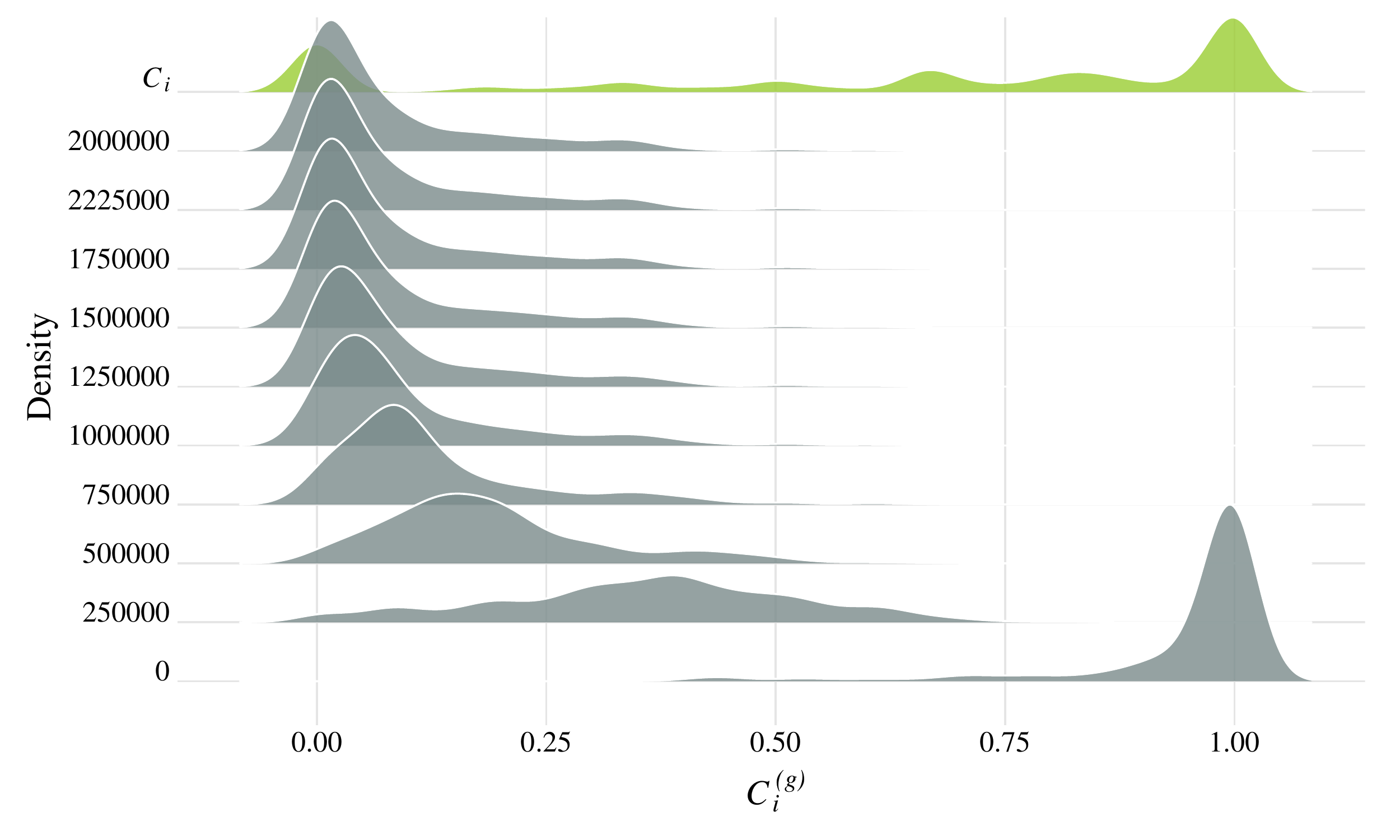}
        %\caption{a}
    \end{minipage}%
     %\ \hspace{5mm} \hspace{5mm}
    \begin{minipage}{0.5\textwidth}
        \centering
        \includegraphics[scale = 0.3, trim = 0cm 0cm 0cm 0cm, clip]{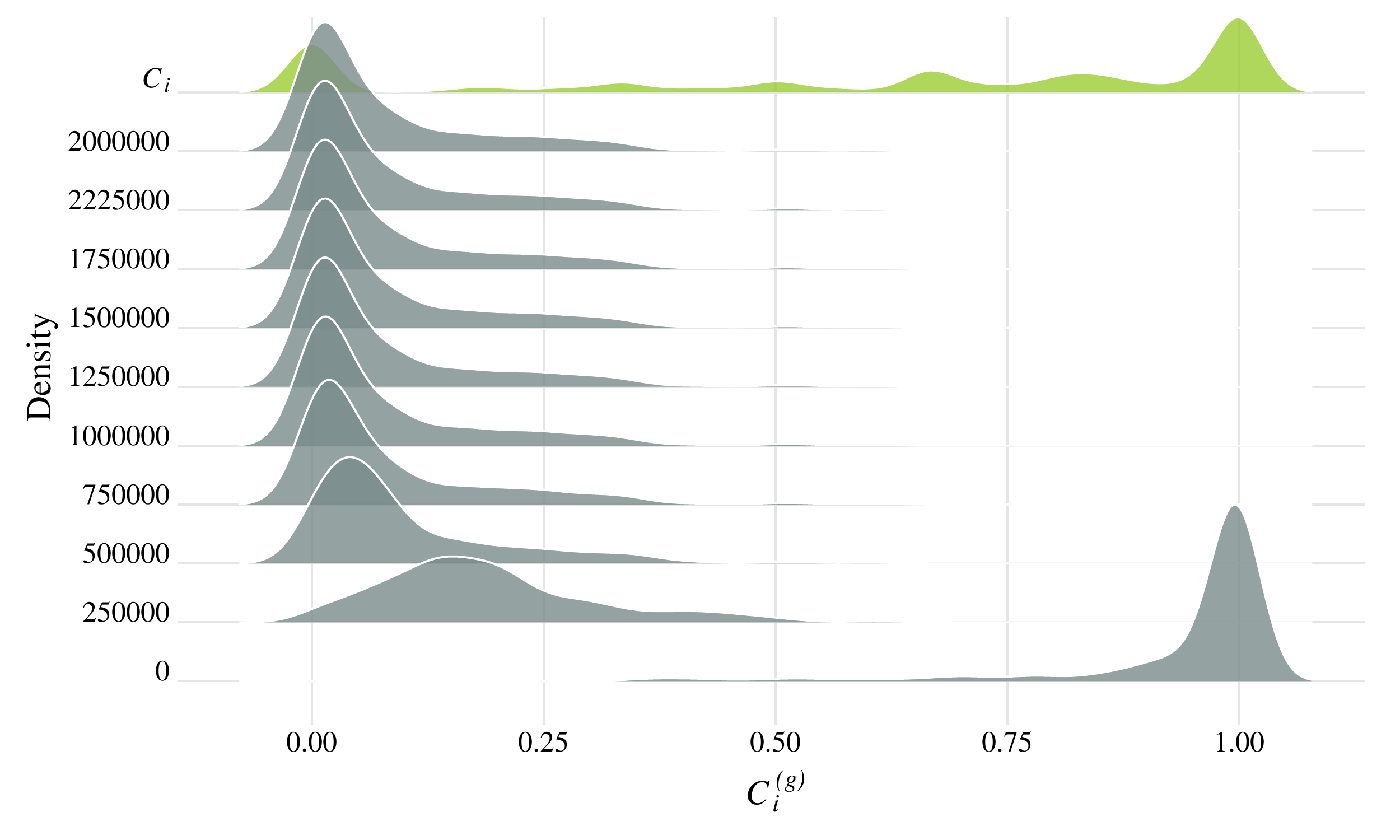}
        %\caption{b}
    \end{minipage}
        \begin{minipage}{0.5\textwidth}
        \centering
        \includegraphics[scale = 0.3, trim = 0cm 0cm 0cm 0cm, clip]{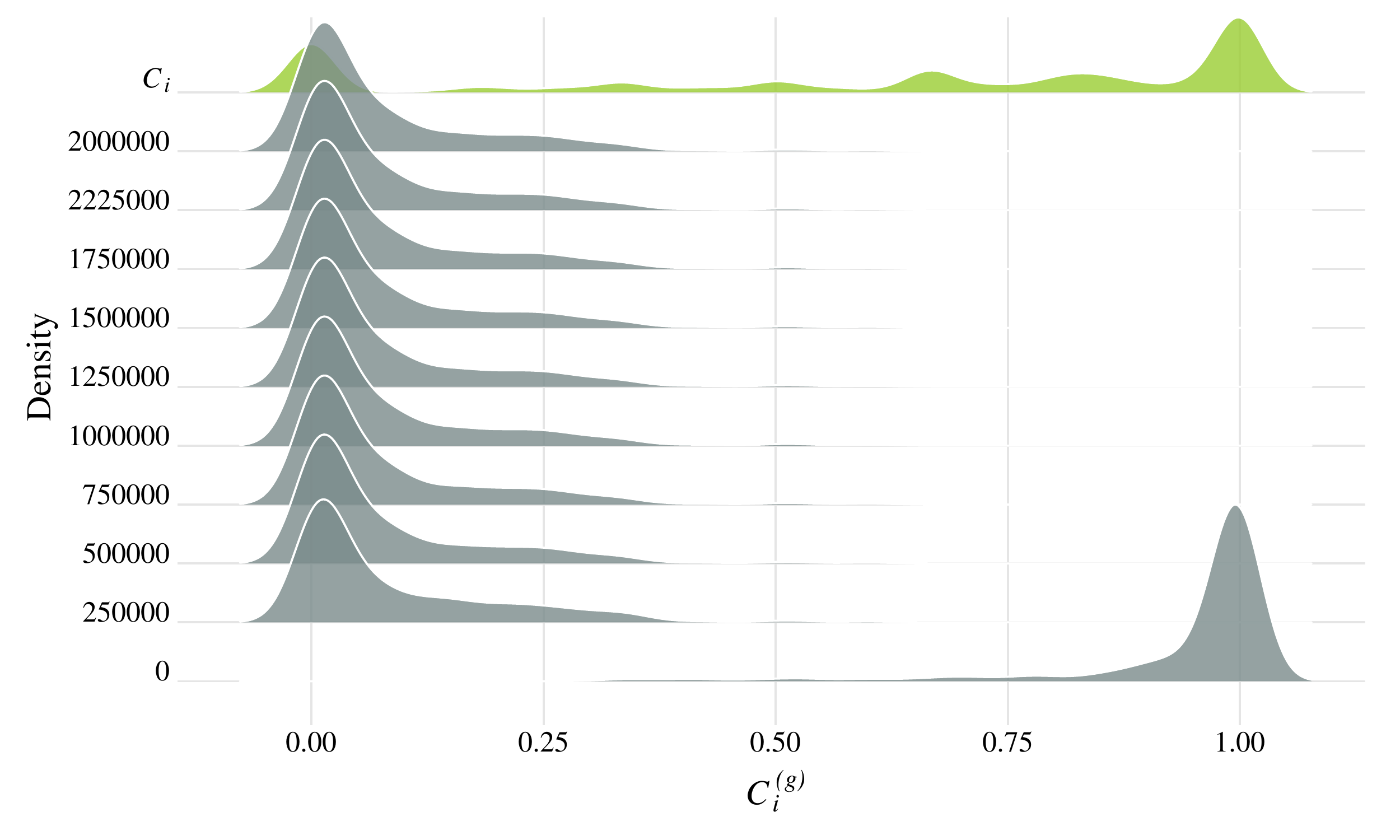}
        %\caption{c}
    \end{minipage}%
     %\ \hspace{5mm} \hspace{5mm}
    \begin{minipage}{0.5\textwidth}
        \centering
        \includegraphics[scale = 0.3, trim = 0cm 0cm 0cm 0cm, clip]{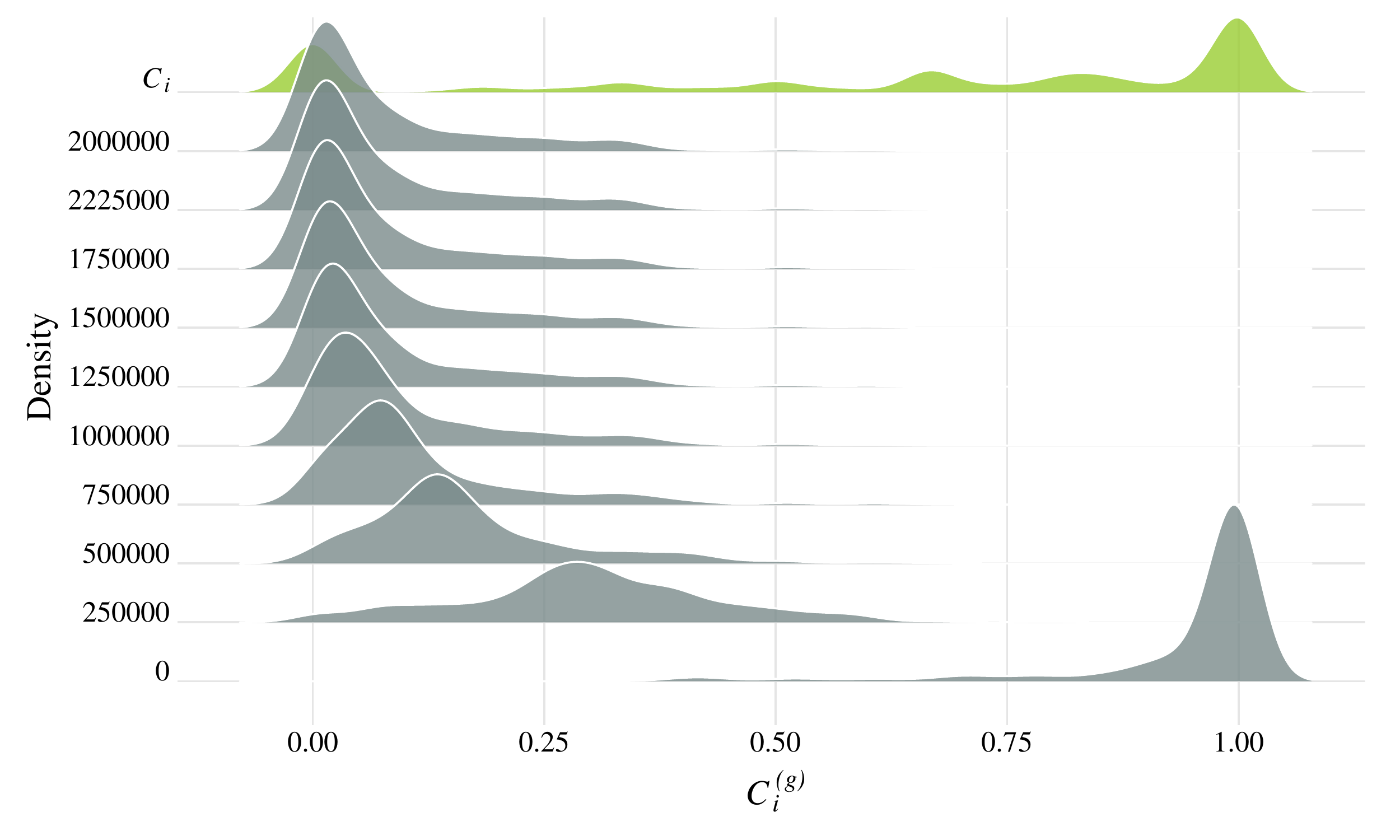}
        %\caption{d}
    \end{minipage}
    \caption{US airports. Density of $C_i^{(g)}$ for different values of $\alpha$ when $\beta = 0$ for cases $F_1$ (upper left), $F_2$ (upper right), $F_3$ (lower left) and $F_4$ (lower right).}
    \label{usaalpha}
\end{figure}

\begin{figure}[!htb]
    \centering
    \begin{minipage}{0.5\textwidth}
        \centering
        \includegraphics[scale = 0.3, trim = 0cm 0cm 0cm 0cm, clip]{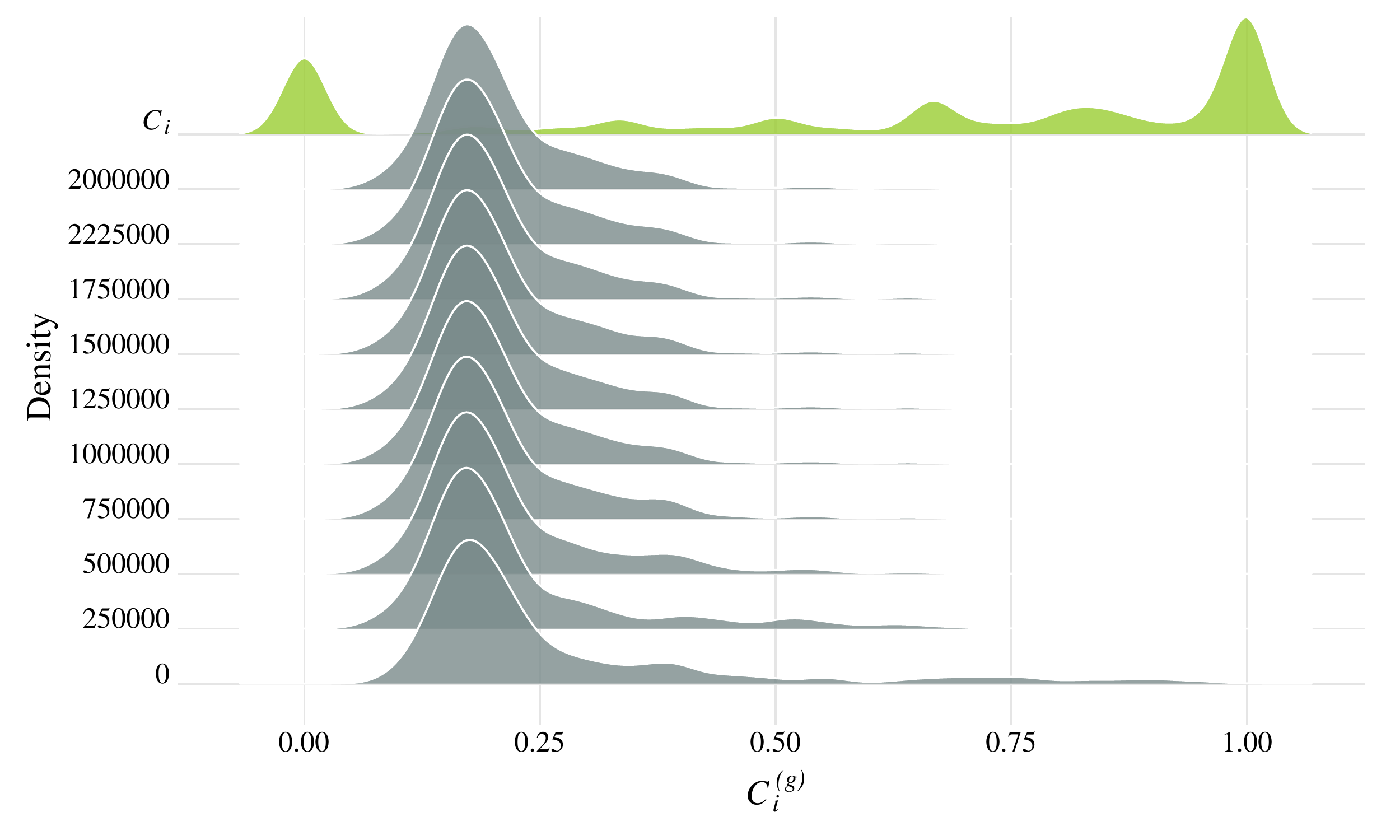}
        %\caption{a}
    \end{minipage}%
     %\ \hspace{5mm} \hspace{5mm}
    \begin{minipage}{0.5\textwidth}
        \centering
        \includegraphics[scale = 0.3, trim = 0cm 0cm 0cm 0cm, clip]{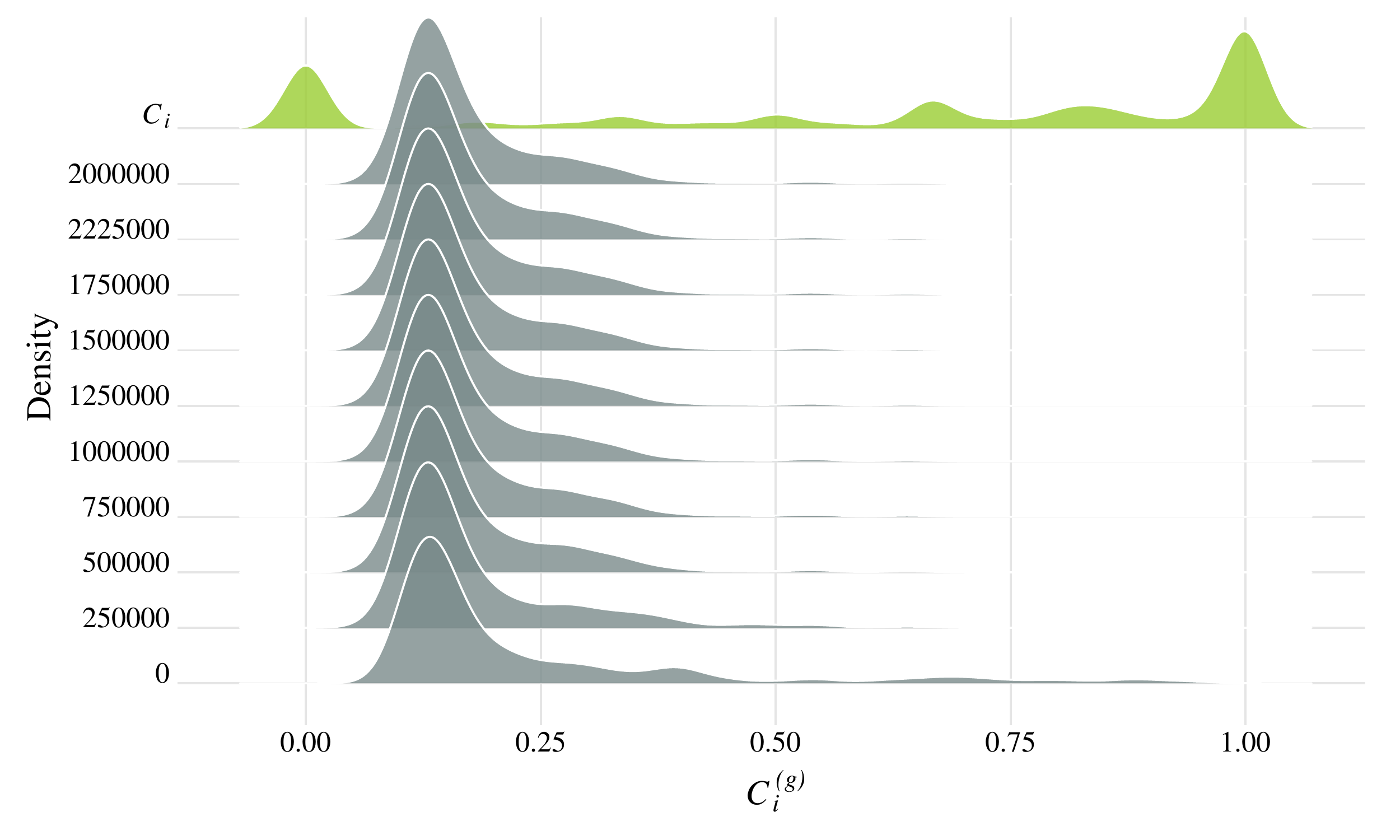}
        %\caption{b}
    \end{minipage}
        \begin{minipage}{0.5\textwidth}
        \centering
        \includegraphics[scale = 0.3, trim = 0cm 0cm 0cm 0cm, clip]{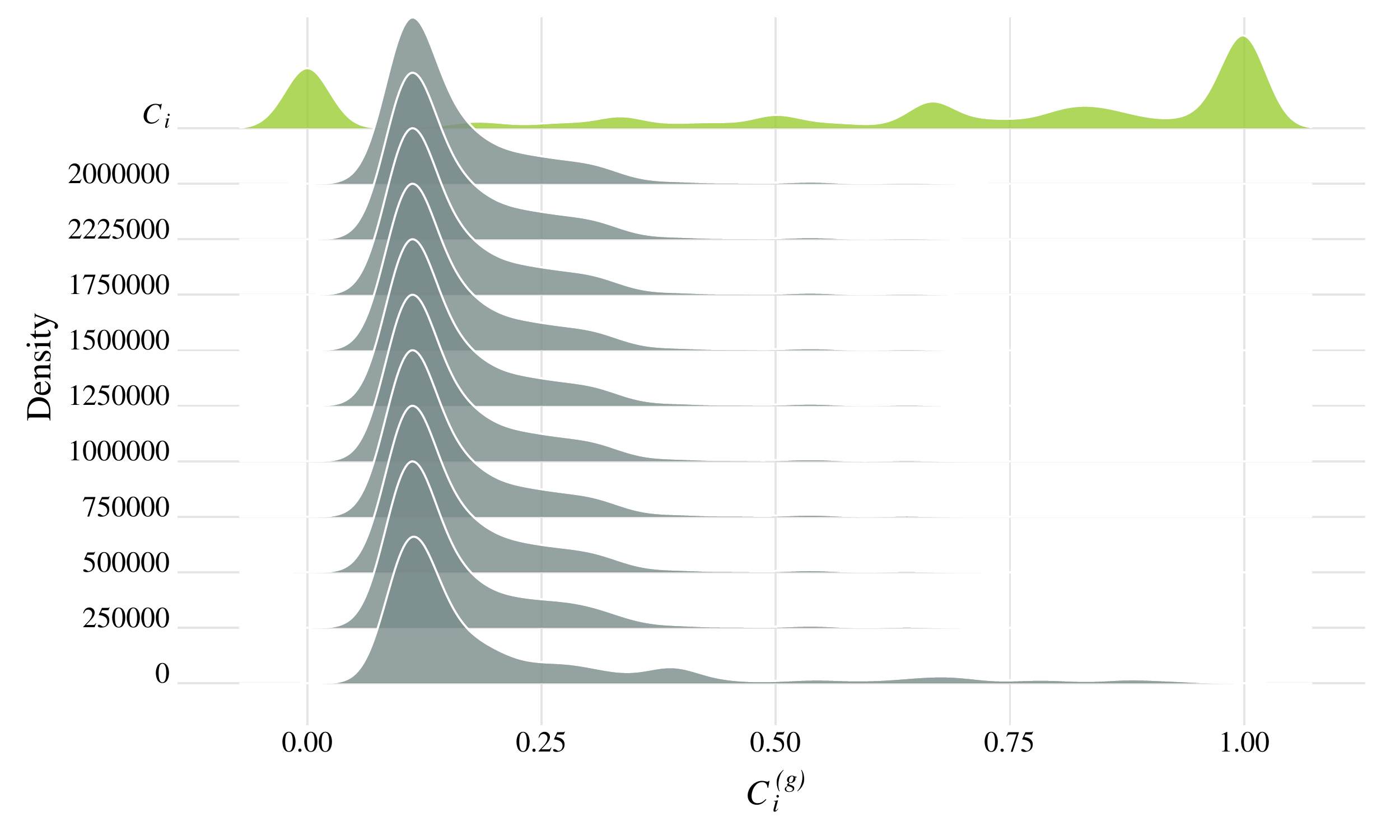}
        %\caption{c}
    \end{minipage}%
     %\ \hspace{5mm} \hspace{5mm}
    \begin{minipage}{0.5\textwidth}
        \centering
        \includegraphics[scale = 0.3, trim = 0cm 0cm 0cm 0cm, clip]{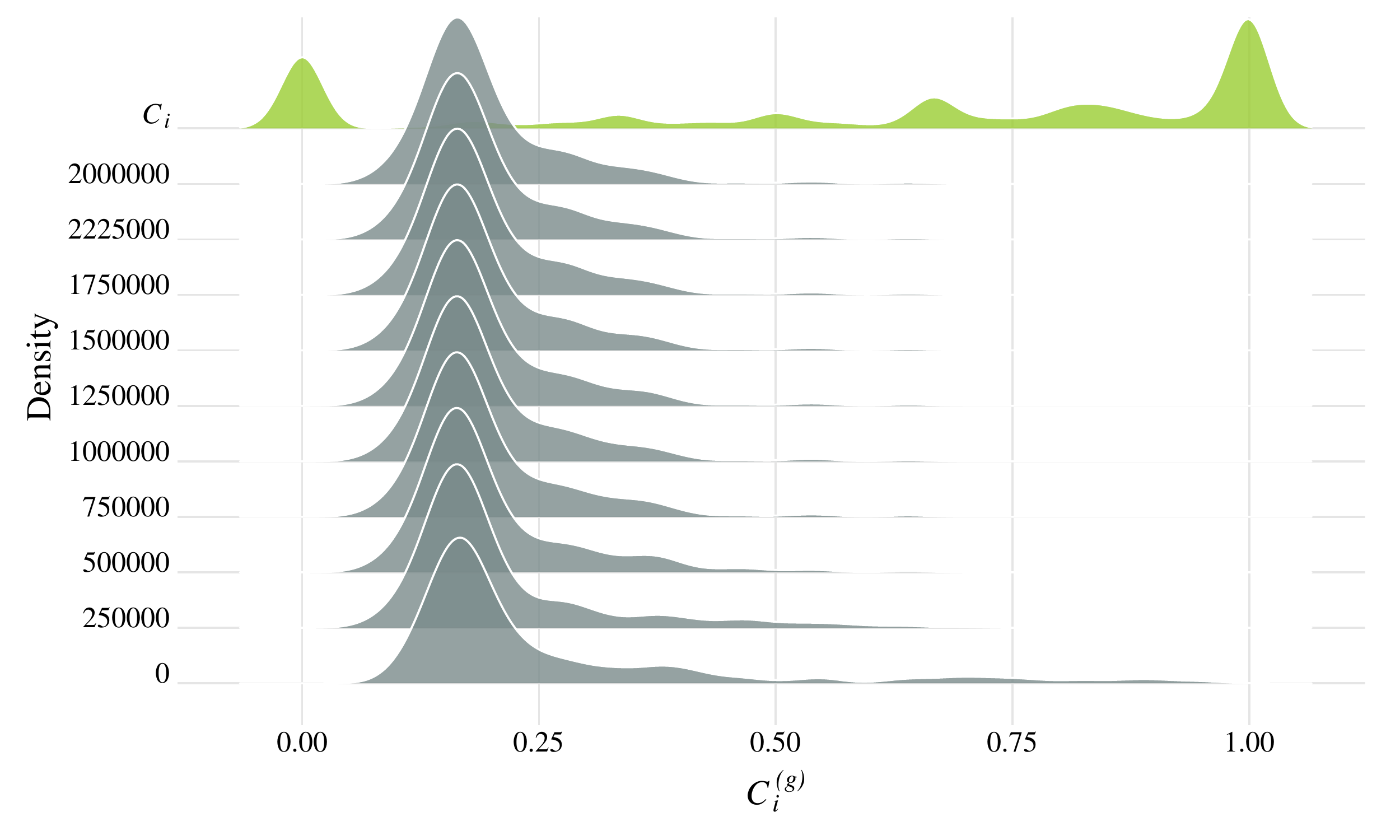}
        %\caption{d}
    \end{minipage}
    \caption{US airports. Density of $C_i^{(g)}$ for different values of $\beta$ when $\alpha = 0$ for cases $F_1$ (upper left), $F_2$ (upper right), $F_3$ (lower left) and $F_4$ (lower right).}
    \label{usabeta}
\end{figure}

In order to study the evolution of the generalized clustering
coefficient $C_i^{(g)}$ when varying $\alpha$ and $\beta$, we
provide a series of diagrams in which, for the network under
examination, the density of the $C_i^{(g)}$ values are reported when
considering fixed values of $\alpha = 0$ or $\beta = 0$  and when
the other thresholds varing.

In particular, for the network under observation, Figure~\ref{usaalpha}
shows different density values for each $\alpha$ when $\beta = 0$,
and Figure~\ref{usabeta} shows each $\beta$ when $\alpha = 0$.
All the figures also report the density values of the local clustering
coefficient $C_i$ (colored in light green).

When $\beta = 0$ (see Figure~\ref{usaalpha})
we can observe the contribution of triples in $T_2$ to
$C_i^{(g)}$. The density of $C_i^{(g)}$ is
more concentrated around the max value $1$ when $\alpha = 0$; however,
when $\alpha$ starts to grow the values shift closer to $0$.

For $\alpha = 0$, Figure~\ref{usabeta} highlights that $C_i^{(g)}$ receives
a small contribution from triples in $T_3$ and the values lay around $0$ as soon
$\beta$ grows.

The density of $C_i$ shows that values are concentrated mainly around $0$ and $1$,
meaning that many airports have a single connection with another airport or have
a strong cohesive structure. When studying $C_i^{(g)}$ it is possible to infer that
some airports with a single connection with a common node have weight profiles
that involve a certain level of interaction for a given threshold. For example,
for low values of $\alpha$ passengers from or to airports $j$
and $k$ often use connection $i$, thus suggesting the establishment of a direct and
intended connection between the two airports.
This is not true when considering triples in $T_3$; here the analysis suggests that a direct
connection among $i$ and $k$ is less useful and passengers still prefer to fly by $j$.

%%%%% fine US airport %%%%%%%

%%%%%%%% inizio celeg %%%%%%%
\subsection{Analysis of the C.elegans network}

The network of nematode Caenorhabditis elegans (C.elegans) has $n=
296$ nodes representing neurons and $m = 1370$ edges occurring when
two neurons are connected by either a synapse or a gap junction; for
each edge,  weights are equal to the number of junctions between
nodes $i$ and $j$. The network has a scale-free organization with
$\gamma \simeq 3.14$~\cite{BA, VCPHC}.

In Figure~\ref{visualization} (right) we show the network visualization,
while Table~\ref{tab_1} reports the basic measures. Note that for
this network we considered the giant component of $296$
nodes while the complete network is composed of $306$ nodes.

In Figure~\ref{strength} (right) we report the strength distributions for
the C.elegans network. Note that the two networks under observation are very
different, especially in the distribution of low and high values of
strength. The weight profiles in Figure~\ref{weight_hist} confirm
such differences, which are mostly caused by a difference of scale in the
values.

\begin{figure}[htbp]
\begin{center}
\includegraphics[scale=.5]{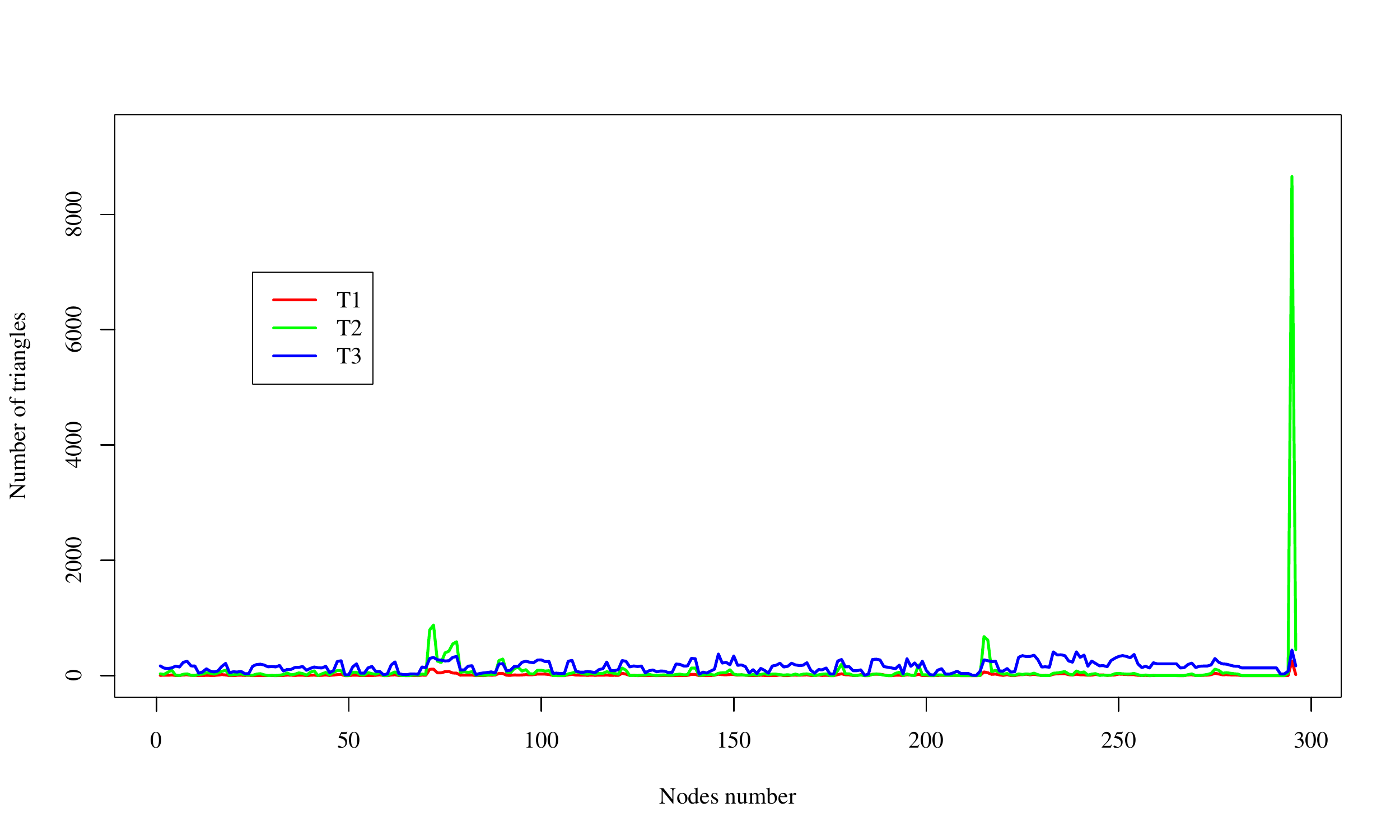}
\caption{C.elegans. Comparison between the number of triangles $T_1$,
the number of potential triangles $T_2$ and the number of
potential triangles $T_3$.} \label{deltatceleg}
\end{center}
\end{figure}

\begin{figure}[htbp]
\begin{center}
\includegraphics[scale=.5]{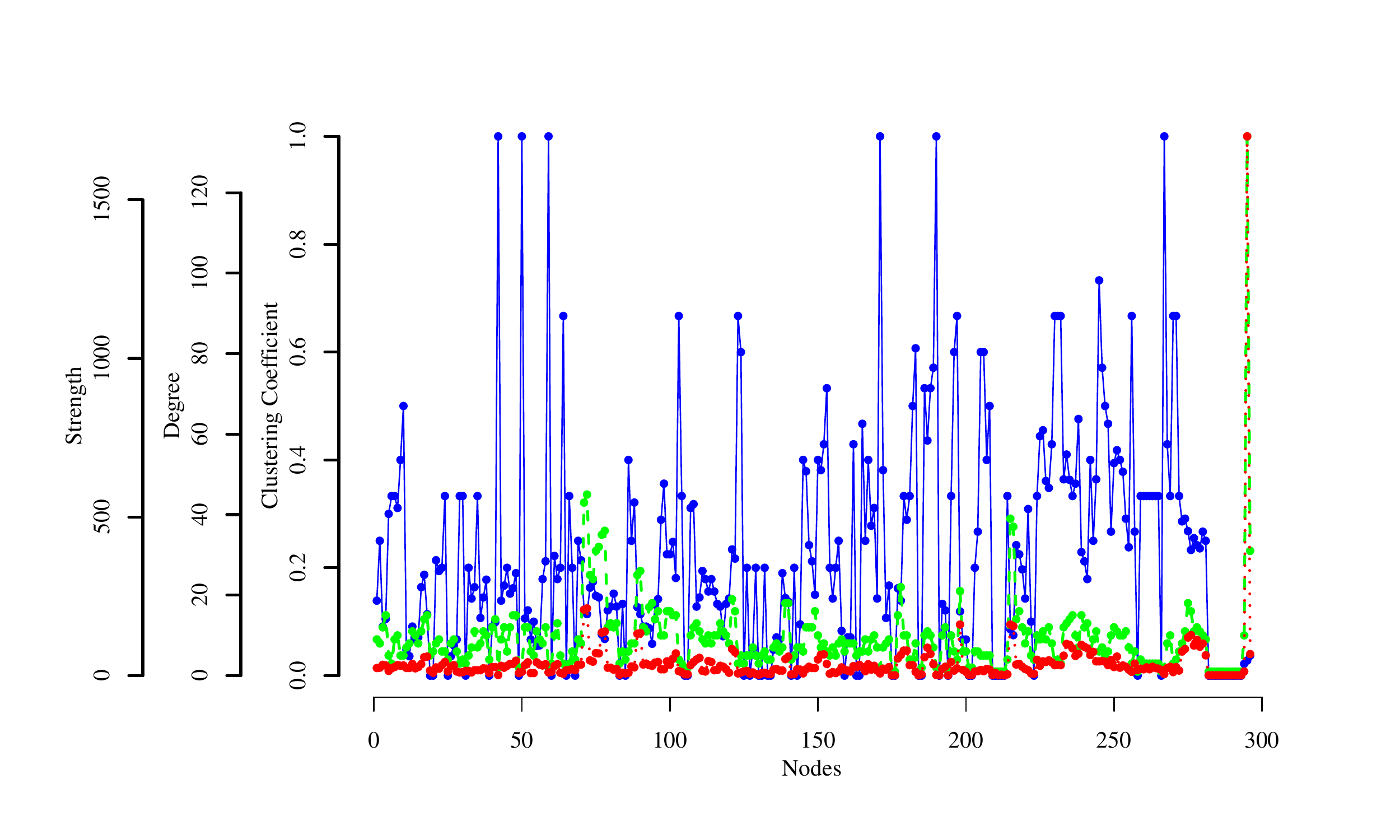}
\caption{C.elegans. Comparison between local clustering coefficient (blue points), degree (red points) and strength (green points).}
\label{CDSceleg}
\end{center}
\end{figure}

The analysis of Figures~\ref{deltatceleg} and~\ref{CDSceleg} depicts a
very different picture for the C.elegans network when compared to the US airports network.
Again, Figure~\ref{deltatceleg} reports the three curves
representing the total number of triangles $|\mathcal{T}_1^{(i)}|$,
the number of potential triples of type
$|\mathcal{T}_2^{(i)}|$ and the number of potential triples
of type $|\mathcal{T}_3^{(i)}|$. Figure~\ref{CDSceleg}
 compares the degree $d_i$ and the local clustering
coefficient $C_i$ for each node $i$. Note that in these benchmark
instances, nodes are enumerated without a particular rule.

In the C.elegans networks, nodes with a higher degree have relatively
small values of local clustering coefficient, whilst nodes with a smaller
degree have, in general, higher values of local clustering coefficient.
This means that small-degree nodes tends to form dense
local neighborhoods, while the neighborhood of hubs is much sparser.
Such observations motivate the limited number of triples in
$T_2$ because, for each node $i$, they are in number of
$\binom{d_i}{2} - |\mathcal{T}_1^{(i)}|$, thus implying that denser neighborhoods
have a smaller number of possible triples.

Note in Figure~\ref{deltatceleg} that node $i = 295$ has a peak
because $|\mathcal{T}_2^{(i)}| = 8658$ when the thresholds $\alpha$
and $\beta$ are null (this is the case of all potential
triples). This is motivated by the fact that its particular neighborhood
is composed of a limited number of triangles in which it is embedded
($|\mathcal{T}_1^{(295)}| = 253$ and $C_{295} = 0.028$) despite its
degree ($d_{295} = 134)$. Choosing two edges on $134$ leads to $8911$
potential triples of type $T_2$ and subtracting $253$
results in $8658$. Such a remarkable presence of triples of type $T_2$
for a single node for the case of $\alpha=\beta=0$ suggests that the
C.elegans network is star-shaped.

Similar arguments can be considered for $T_3$; indeed, we have a
small number of potential triples for both small-degree
nodes and hubs, since low values of degree allow for a smaller amount
of transitive closure.

\begin{figure}[!htb]
    \centering
    \begin{minipage}{0.5\textwidth}
        \centering
        \includegraphics[scale = 0.4, trim = 5cm 0cm 5cm 0cm, clip]{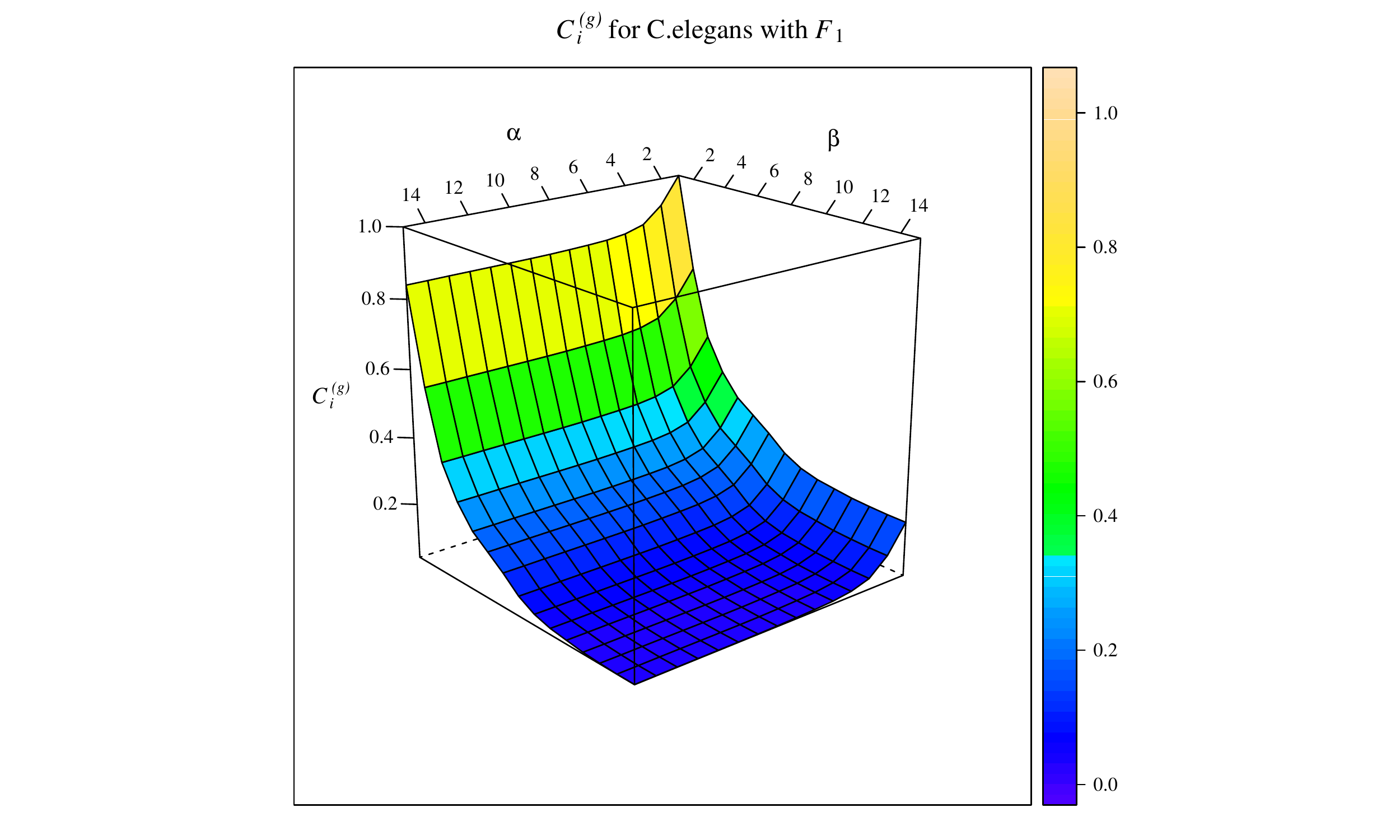}
        %\caption{a}
    \end{minipage}%
     %\ \hspace{5mm} \hspace{5mm}
    \begin{minipage}{0.5\textwidth}
        \centering
        \includegraphics[scale = 0.4, trim = 5cm 0cm 5cm 0cm, clip]{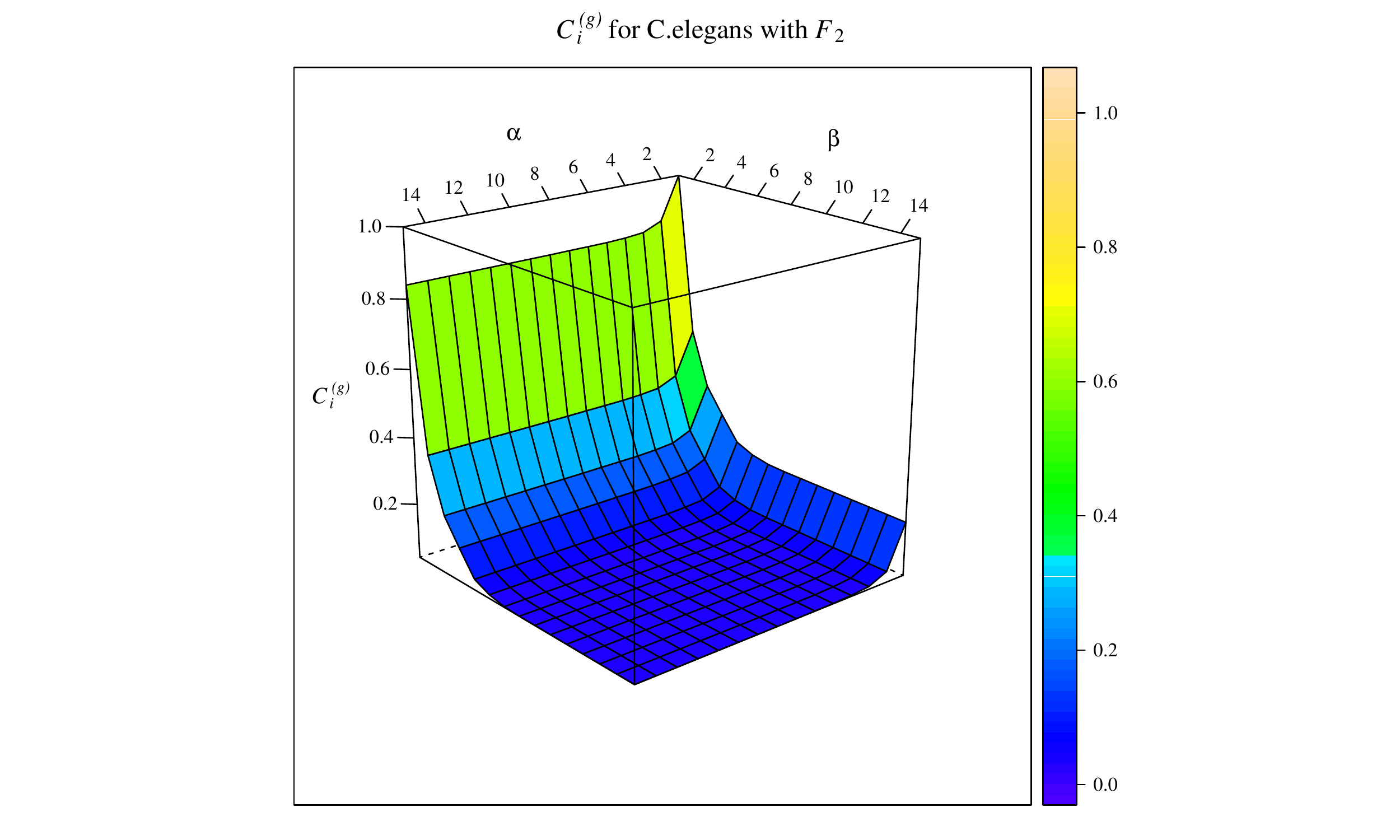}
        %\caption{b}
    \end{minipage}
        \begin{minipage}{0.5\textwidth}
        \centering
        \includegraphics[scale = 0.4, trim = 5cm 0cm 5cm 0cm, clip]{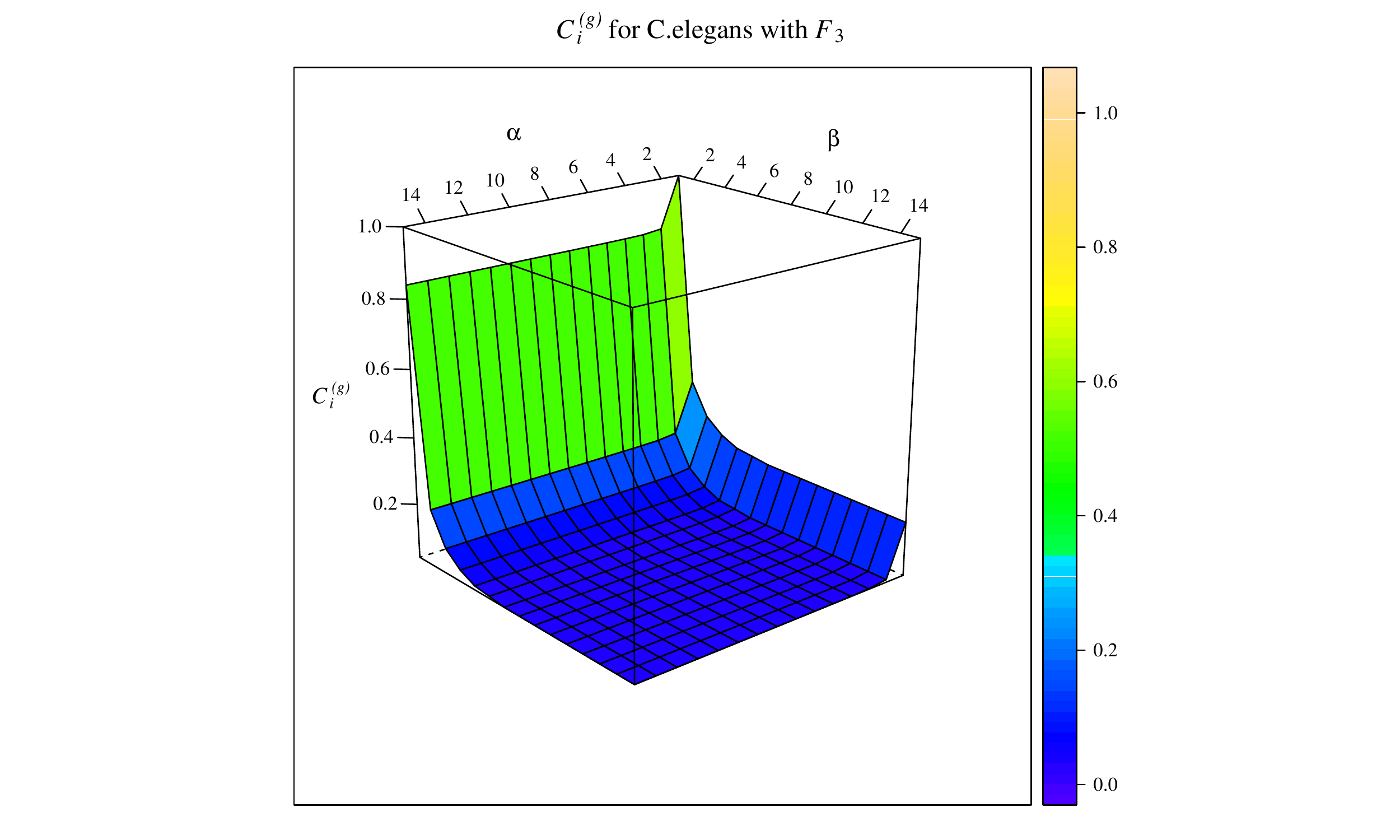}
        %\caption{c}
    \end{minipage}%
     %\ \hspace{5mm} \hspace{5mm}
    \begin{minipage}{0.5\textwidth}
        \centering
        \includegraphics[scale = 0.4, trim = 5cm 0cm 5cm 0cm, clip]{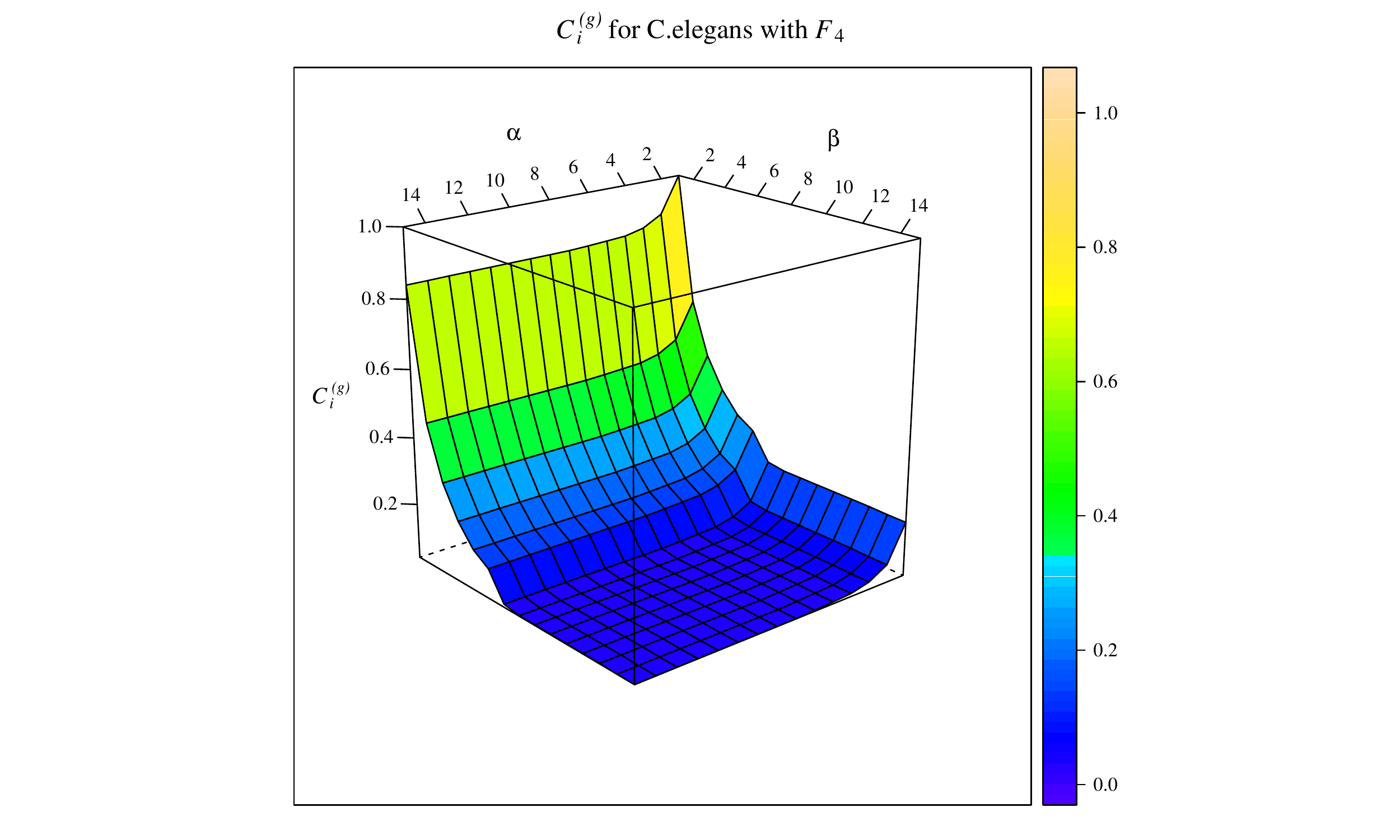}
        %\caption{d}
    \end{minipage}
    \caption{C.elegans. Average values of $C_i^{(g)}$ for cases $F_1$ (upper left), $F_2$ (upper right), $F_3$ (lower left) and $F_4$ (lower right).}
    \label{celegCg}
\end{figure}

Figure~\ref{celegCg} reports the same plots for the C.elegans network
and same comments on the general behavior can be repeated as for the
previous network.
The main difference is the gentler slope, which occurs due to
 the profile of weight distribution being less concentrated on
lower values when compared to the US airports network (see also Figure~\ref{weight_hist}).

\begin{figure}[!htb]
    \centering
    \begin{minipage}{0.5\textwidth}
        \centering
        \includegraphics[scale = 0.3, trim = 0cm 0cm 0cm 0cm, clip]{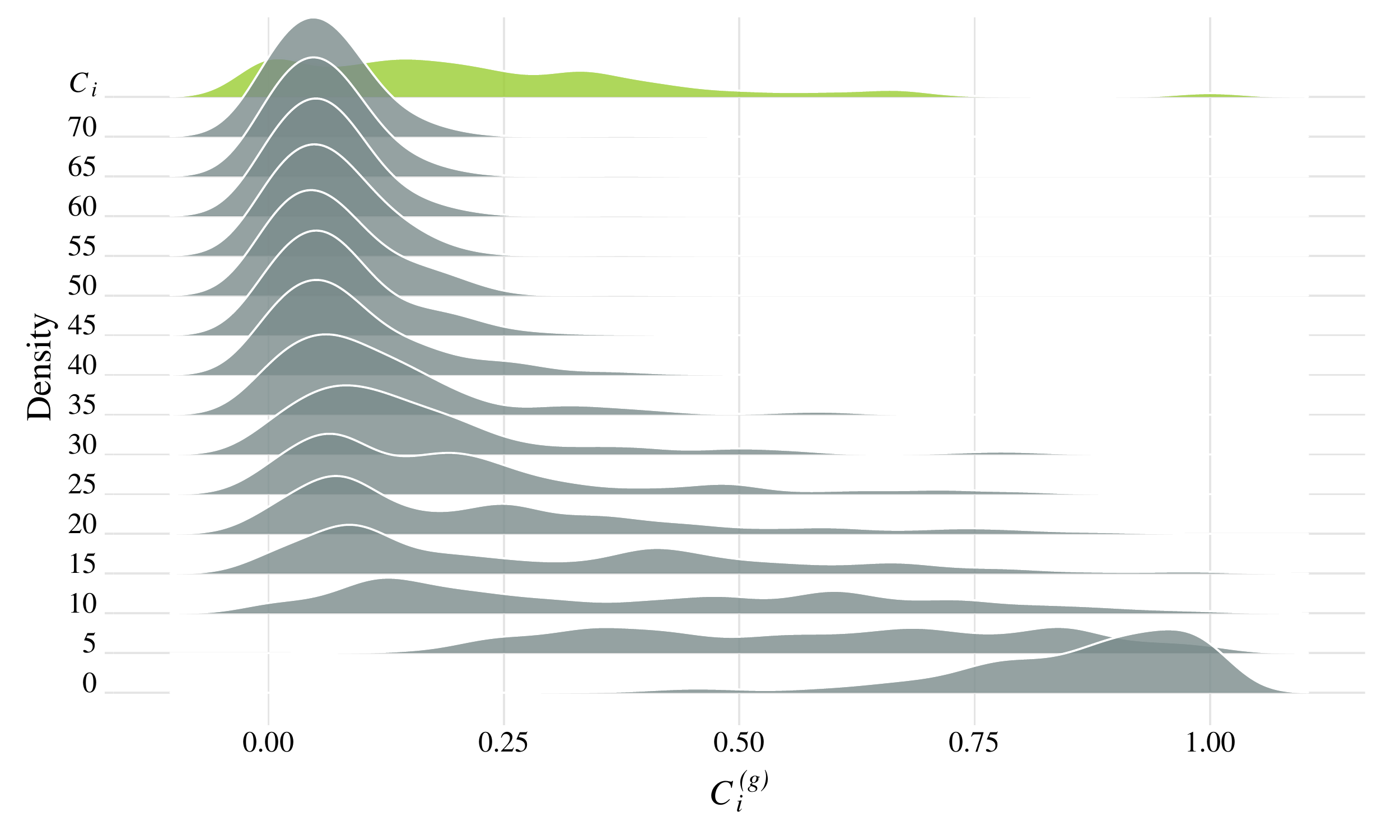}
        %\caption{a}
    \end{minipage}%
     %\ \hspace{5mm} \hspace{5mm}
    \begin{minipage}{0.5\textwidth}
        \centering
        \includegraphics[scale = 0.3, trim = 0cm 0cm 0cm 0cm, clip]{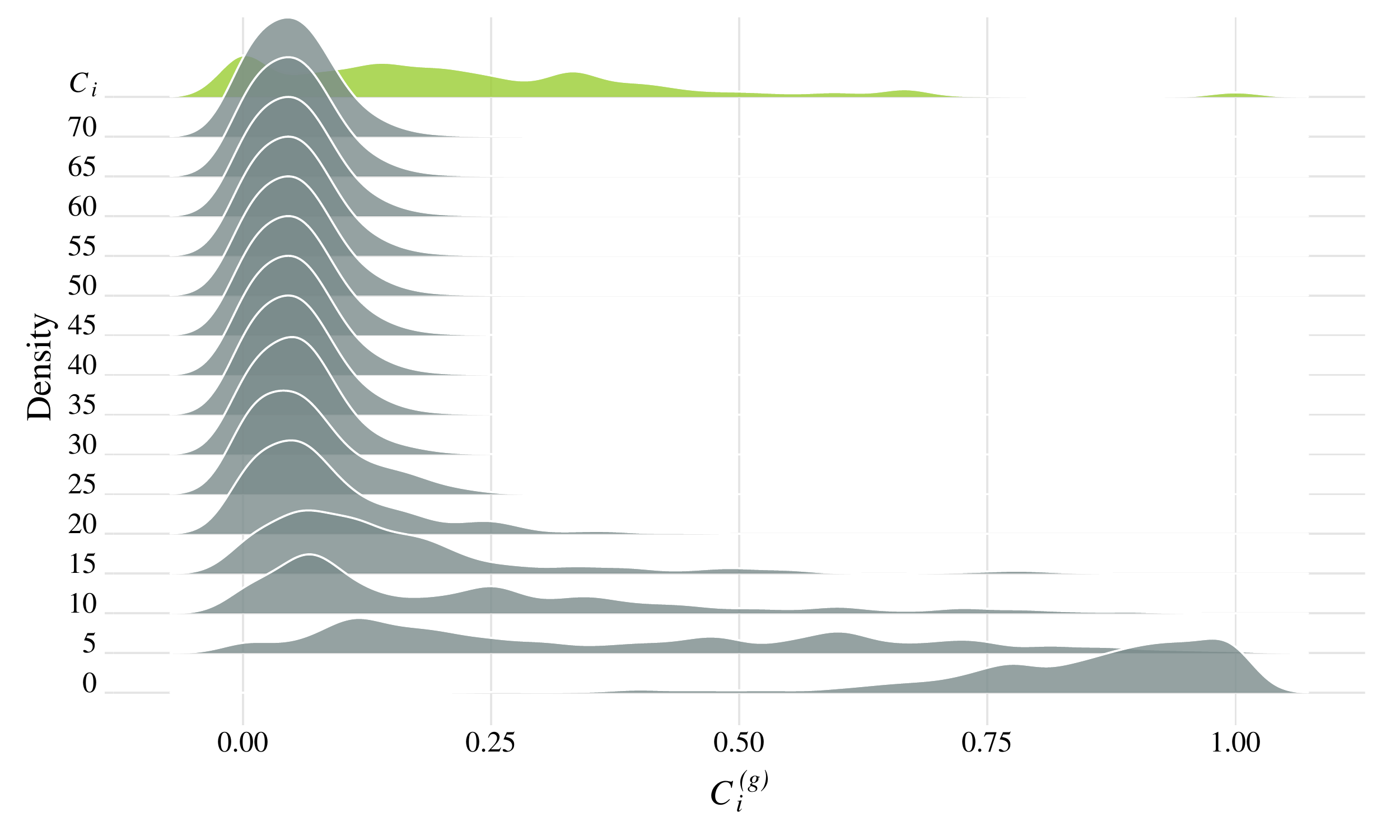}
        %\caption{b}
    \end{minipage}
        \begin{minipage}{0.5\textwidth}
        \centering
        \includegraphics[scale = 0.3, trim = 0cm 0cm 0cm 0cm, clip]{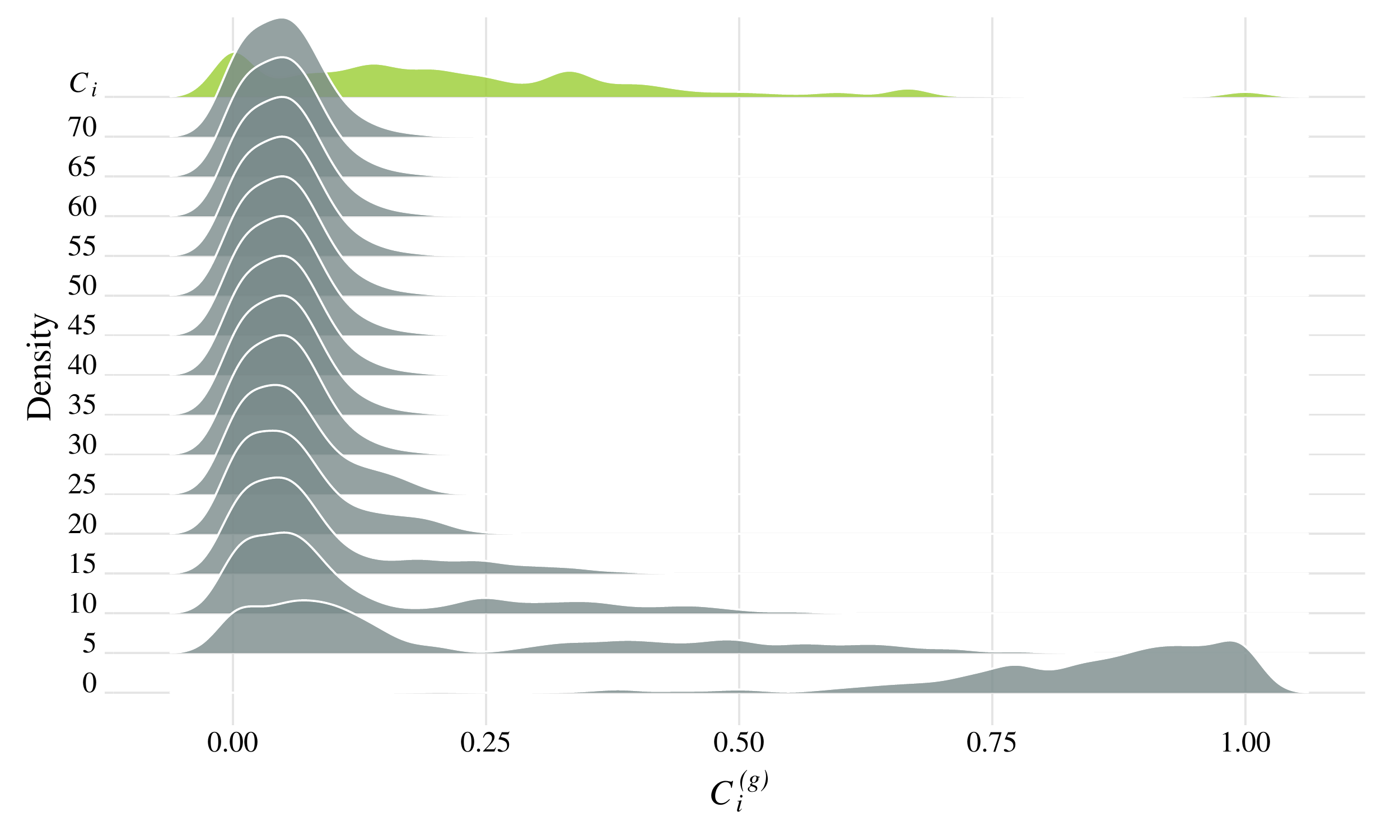}
        %\caption{c}
    \end{minipage}%
     %\ \hspace{5mm} \hspace{5mm}
    \begin{minipage}{0.5\textwidth}
        \centering
        \includegraphics[scale = 0.3, trim = 0cm 0cm 0cm 0cm, clip]{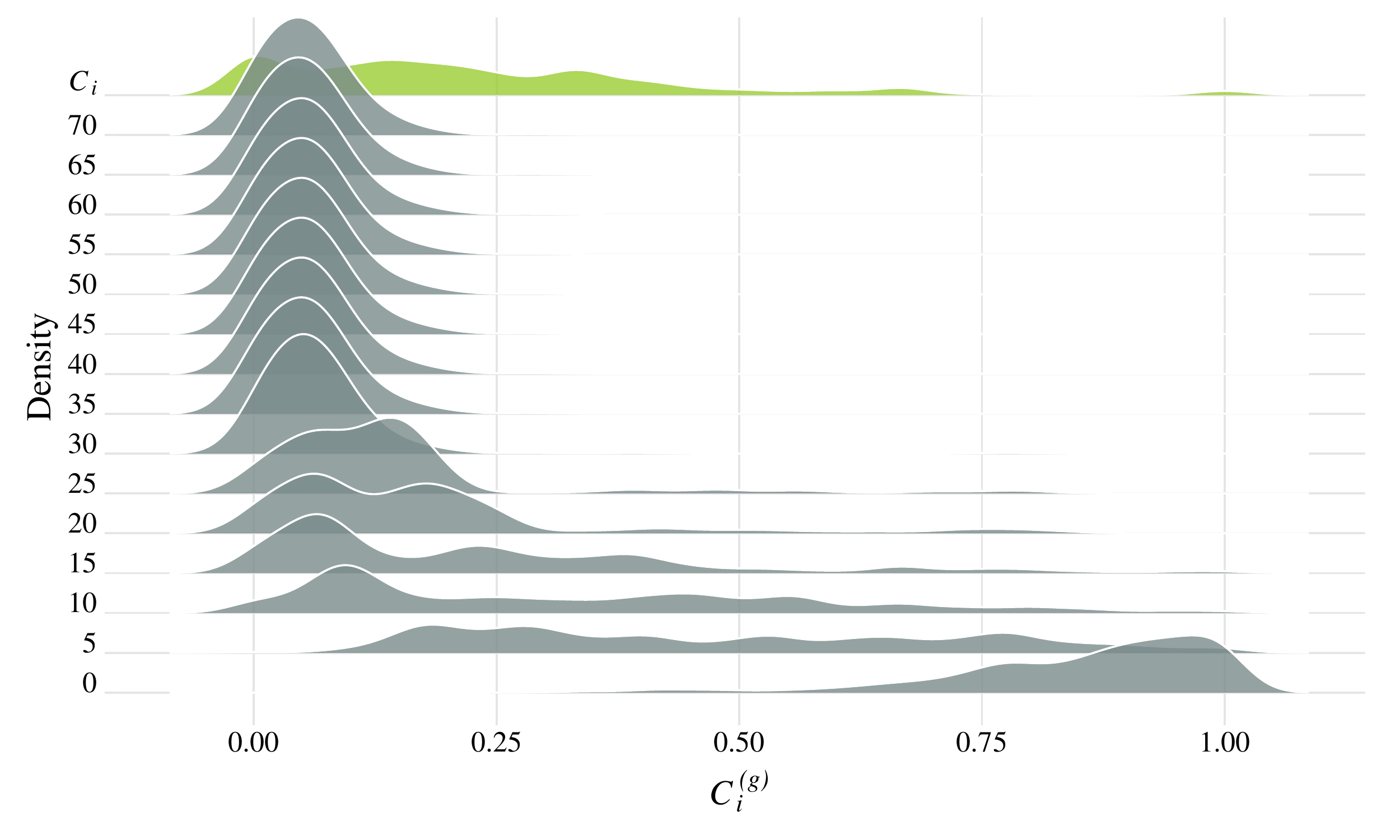}
        %\caption{d}
    \end{minipage}
    \caption{C.elegans. Density of $C_i^{(g)}$ for different values of $\alpha$ when $\beta = 0$ for cases $F_1$ (upper left), $F_2$ (upper right), $F_3$ (lower left) and $F_4$ (lower right).}
    \label{celegalpha}
\end{figure}

\begin{figure}[!htb]
    \centering
    \begin{minipage}{0.5\textwidth}
        \centering
        \includegraphics[scale = 0.3, trim = 0cm 0cm 0cm 0cm, clip]{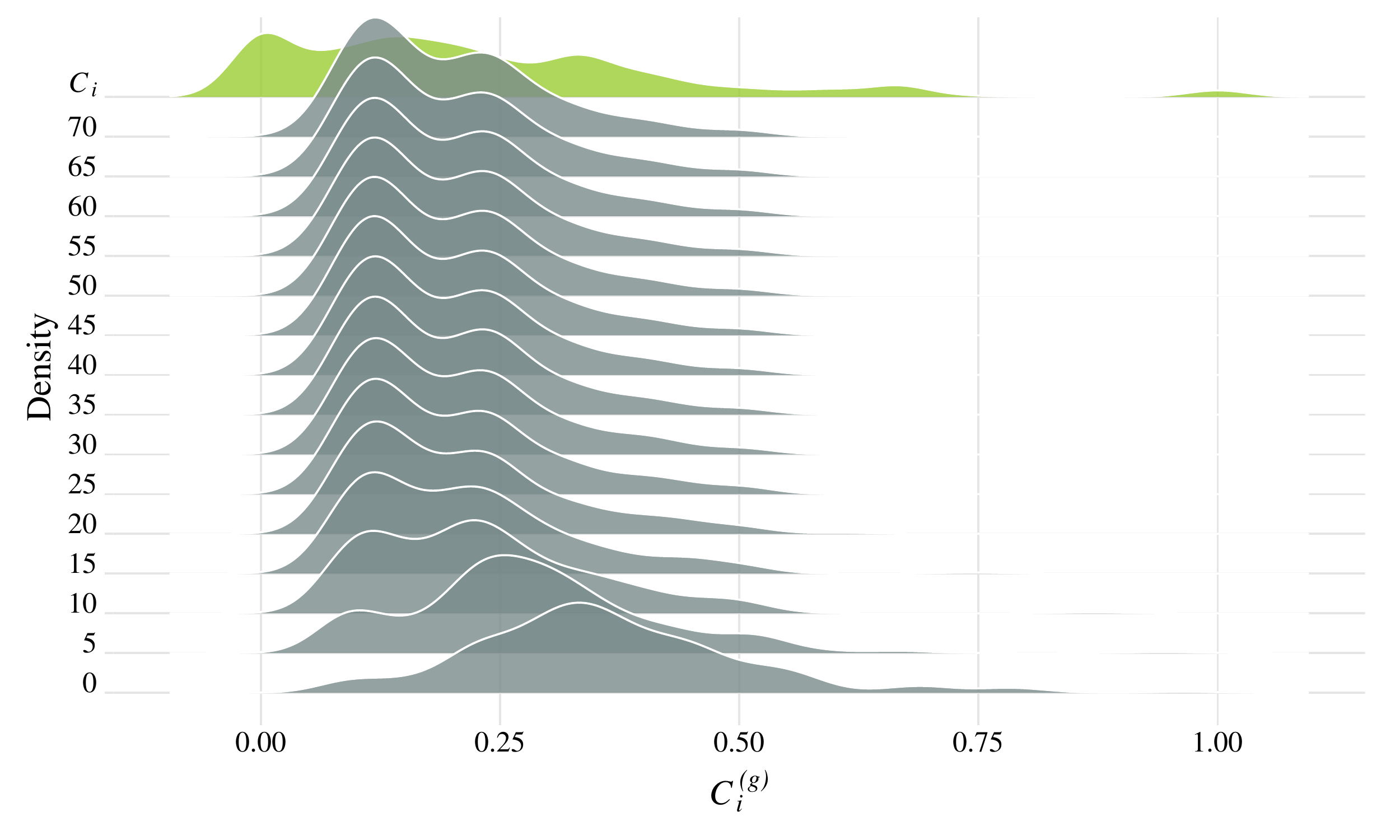}
        %\caption{a}
    \end{minipage}%
     %\ \hspace{5mm} \hspace{5mm}
    \begin{minipage}{0.5\textwidth}
        \centering
        \includegraphics[scale = 0.3, trim = 0cm 0cm 0cm 0cm, clip]{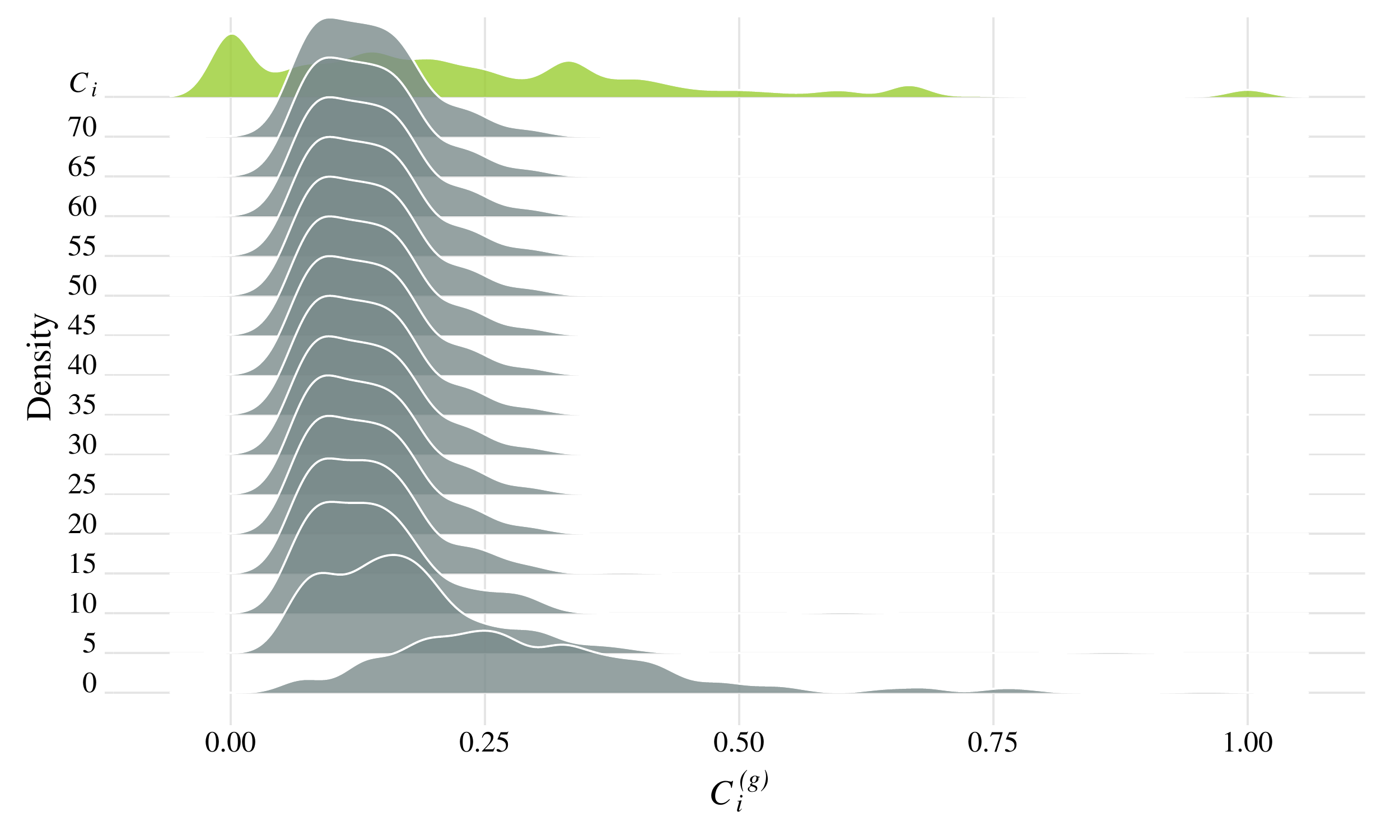}
        %\caption{b}
    \end{minipage}
        \begin{minipage}{0.5\textwidth}
        \centering
        \includegraphics[scale = 0.3, trim = 0cm 0cm 0cm 0cm, clip]{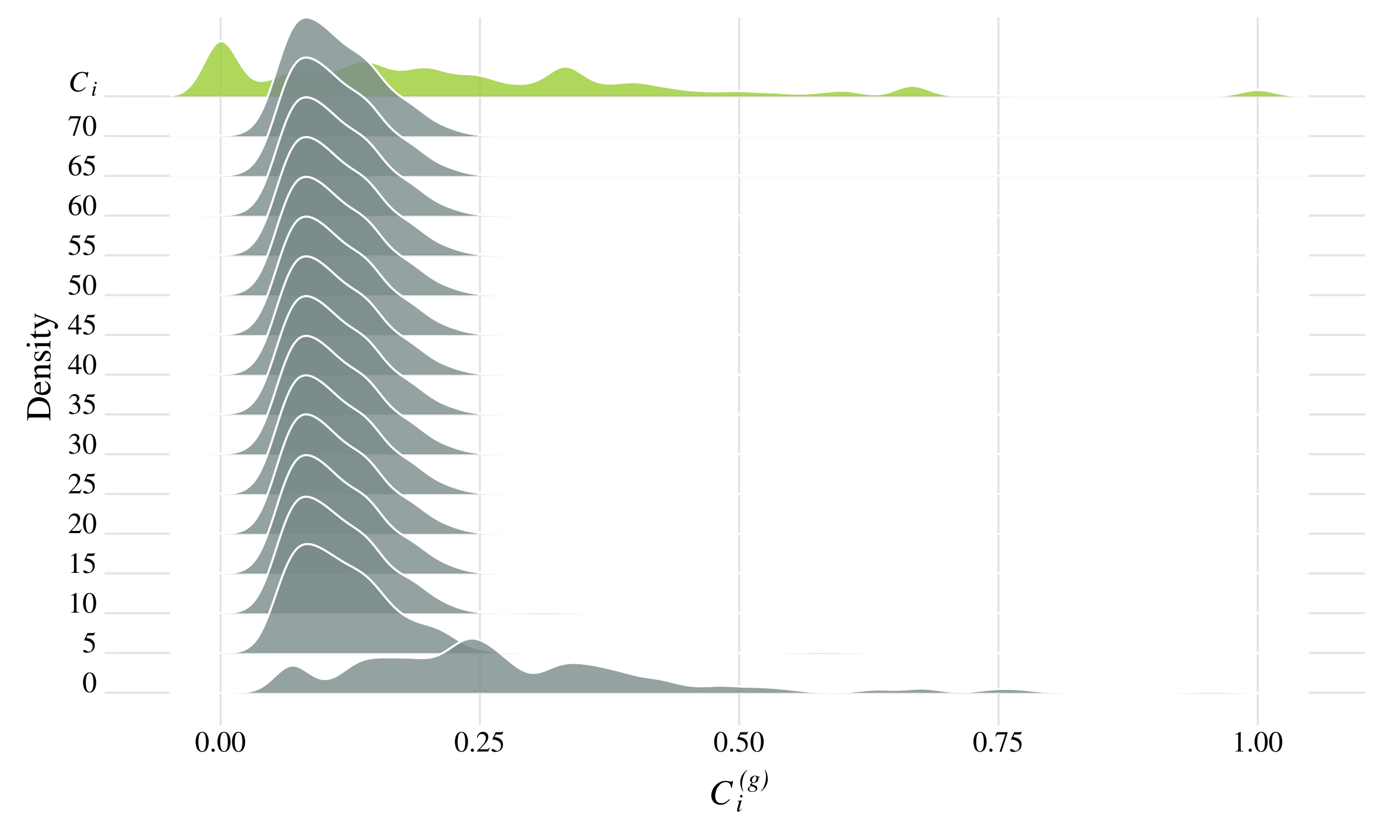}
        %\caption{c}
    \end{minipage}%
     %\ \hspace{5mm} \hspace{5mm}
    \begin{minipage}{0.5\textwidth}
        \centering
        \includegraphics[scale = 0.3, trim = 0cm 0cm 0cm 0cm, clip]{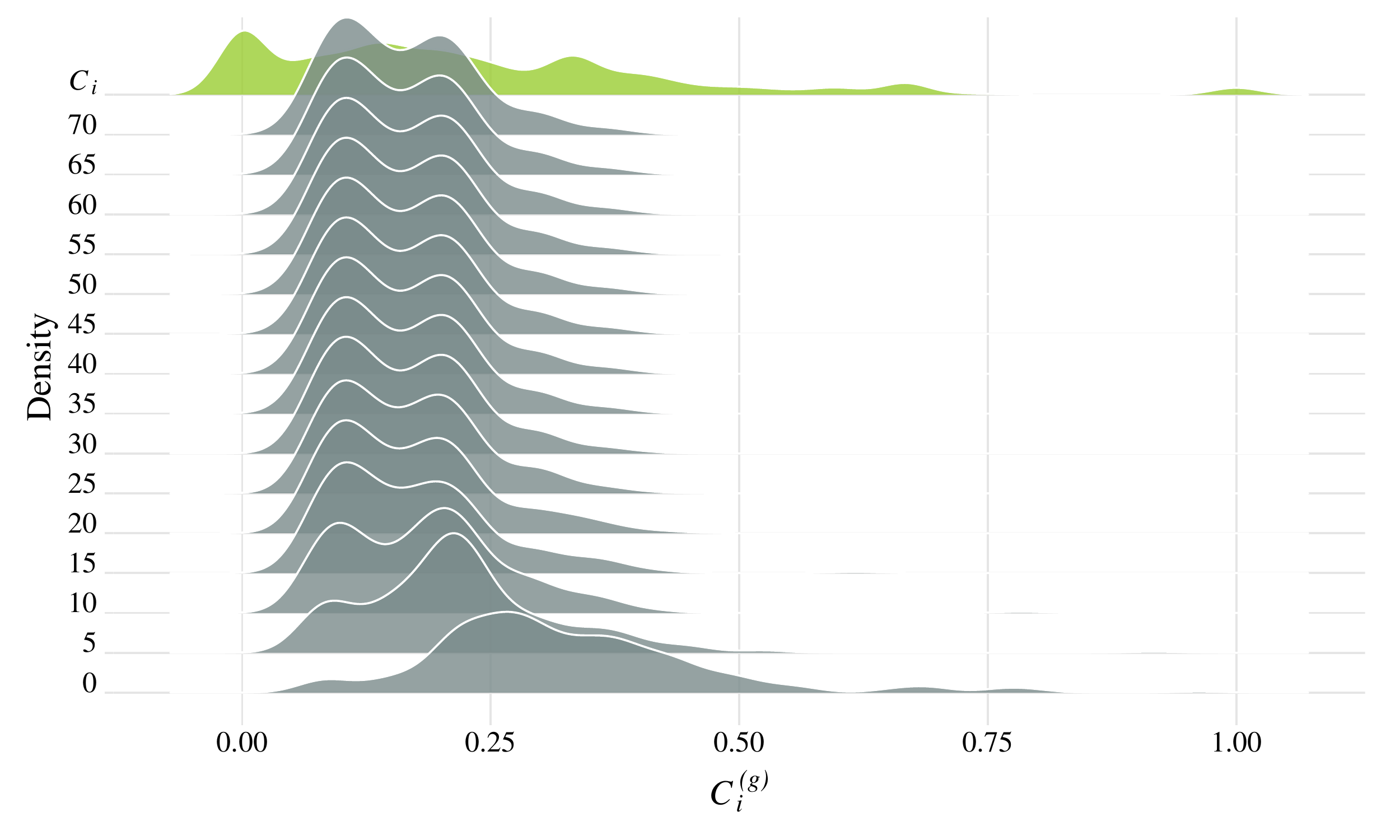}
        %\caption{d}
    \end{minipage}
    \caption{C.elegans. Density of $C_i^{(g)}$ for different values of $\beta$ when $\alpha = 0$ for cases $F_1$ (upper left), $F_2$ (upper right), $F_3$ (lower left) and $F_4$ (lower right).}
    \label{celegbeta}
\end{figure}

The analysis of the C.elegans network is completed with the series of diagrams in which
the density of the $C_i^{(g)}$ values are reported when considering fixed values
of $\alpha = 0$ or $\beta = 0$ and varying the other threshold.

Even for this network, Figure~\ref{celegalpha}
shows different density values for each $\alpha$ when $\beta = 0$
and Figure~\ref{celegbeta} for each $\beta$ when $\alpha = 0$. Note that
all the figures report the density values of the local clustering
coefficient $C_i$ (colored in light green).

When $\beta = 0$ (see Figure~\ref{celegalpha}) the contribution of triples
in $T_2$ makes the density of $C_i^{(g)}$
more concentrated around the max value $1$ when $\alpha = 0$;
when $\alpha$ starts to grow the values shift closer to $0$.
Such an effect is present in both the networks under observation but it is more evident for the
C.elegans.

Similarly, at the US airports network, for $\alpha = 0$, Figure~\ref{celegbeta} highlights
that $C_i^{(g)}$ receives a small contribution from
triples in $T_3$ and the values lay around $0$ as soon as
$\beta$ grows.

The observation of the way in which $C_i$ density lays seems to
affirm that the network has a small cohesive structure, since the
values are mostly around $0$ or on low values ($< 0.3$). The study
of $C_i^{(g)}$ values highlights that for small values of $\alpha$
there exists an intense interaction among alters, i.e. many neurons
undergo a certain level of mutual influence when connected to a
common neuron. When considering triples in $T_3$ and in particular
for $F_1$, the density remains away from $0$, i.e. transitive
influence is always present even for growing values of $\beta$,
mostly because of the very high strength of node $295$.

The different figures confirm that, for the two networks under
observation, the main contribution to $C_i^{(g)}$ is provided by the
triangles in $T_2$, i.e. their structures and weight profiles
cause the networks to be more prone to close triples in
$T_2$ rather than in $T_3$.

%%%%%%%% fine Celeg %%%%%%%%%%%%

%%%%%%%%%%%%%%%%%%%%%%
\section{Conclusions and future research lines}\label{conclusioni}

In complex systems, the way in which members behave is influenced by their interactions with one another,
as well as by other, not always explicit, phenomena.
Networks are a special case in which
interactions can be studied in more formal ways. In this regards, certain aspects of the
network structure, for instance, the neighborhood around a node or different ways of clustering,
allow one to study important characteristics as local or global cohesive groups.

A classical measure used to study the local cohesiveness is the cluster coefficient, which has been
used in almost every network analysis. However, when a weighted network is considered,
such a measure starts to become ambiguous since all the introduced measures are sensitive
to the degree, as well as the strength profiles, of a node.

Despite the classical clustering coefficient being defined as a measure of the combinatoric structure of the network,
it does not have any ability to provide information when links, rather than being established,
are indirectly induced
by strong cooperations among the formally linked nodes. This occurs when two alters of a common ego have an increased likelihood of meeting
due to the fact that the social motivations are strong enough or that the weight between an alter of its alters has
such an intensity that a connection with the ego is admissible.

This paper deals with a novel definition of the clustering
coefficient for weighted networks in that triangles are viewed under such
social perspectives, thus allowing consideration of cases whereby one of the
edges is missing between three nodes. The propensity to induce missing edges
is studied by means of two thresholds $\alpha$ and $\beta$, which capture key information
on the strength profile of a node's neighborhood.

The definition of two types of triangles, $T_2$ and $T_3$, serves two purposes: on the one hand, they
model the evidence that transitive relations among the nodes appear
when the existing links are strong enough; on the other hand, an understanding
of the number and types of the triangles around the nodes when
$\alpha=\beta=0$ identify equivalent classes of networks on the
basis of their topological structures.

The experiments on two real networks, with many different peculiar
characteristics, highlight the ability of the proposed measure to
express the hidden influences between nodes according to the weight
profiles. A thorough computational exercise has also shown the
sensitivity of the networks to the thresholds values, thus allowing
us to obtain further information.

Future research should be devoted in order to extend this approach
to more complicated problems. For instance, the topological structure of the network can
be discussed in more details. In this respect, note that one can
introduce a novel formulation of the concepts of hubs and centrality
measures on the basis of the social connections among the nodes,
according to our definition of induced indirect links. In this
context, one is able to generalize the exploration to the cases when
$\alpha$ and $\beta $  are not necessarily equal to zero.

\end{document}